\documentclass[10pt,aps,prd,twocolumn,amsmath,amssymb,nofootinbib,eqsecnum,superscriptaddress]{revtex4-2}

\usepackage{mathtools}					%% mathclap, etc.
\usepackage{centernot}                  %% slashed delta
\usepackage{xcolor}			            %% \color
\usepackage{enumitem}                   %% enumerate
\usepackage{textcomp}                   %% textquotesingle 

\usepackage[colorlinks=true,
            citecolor=red,
            linkcolor=blue,
            urlcolor=violet,
            filecolor=cyan,
            backref=false]{hyperref}

\newcommand\bs[1]{\boldsymbol{#1}}

\newcommand\dd{\mathrm{d}}
\newcommand\di{\mathrm{div}}

\newcommand\feq{\mathrel{\phantom{=}}}

\renewcommand{\Re}{\operatorname{Re}}
\renewcommand{\Im}{\operatorname{Im}}

\newcommand{\lie}{\pounds}
\newcommand{\erf}{\operatorname{erf}}
\newcommand{\gd}{\operatorname{gd}}
\newcommand{\csch}{\operatorname{csch}}
\newcommand{\sgn}{\operatorname{sgn}}

\newcommand{\E}{\operatorname{E}}
\newcommand{\T}{\operatorname{T}}
\newcommand{\U}{\operatorname{U}}

\begin{document}

%%%%%%%%%%%%%%%%%%%%%%%%%%%%%%%%%%%%%%%%%%%%%%%%%%%%%%%%%%%%%%%%%%%%%%%%%%%%%%%%%%%%%%
%%%%%%%%%%%%%%%%%%%%%%%%%%%%%%%%%%%%%%%%%%%%%%%%%%%%%%%%%%%%%%%%%%%%%%%%%%%%%%%%%%%%%%
%% TITLE

\title{Non-local scalar fields in static spacetimes via heat kernels}

\author{Ivan Kol\'a\v{r}}
\email{i.kolar@rug.nl}
\affiliation{Van Swinderen Institute, University of Groningen, 9747 AG, Groningen, Netherlands}

\date{\today}

\begin{abstract}
We solve the non-local equation ${-e^{-\ell^2\square}\square{\phi}=J}$ for i) static scalar fields in static spacetimes and ii) time-dependent scalar fields in ultrastatic spacetimes. Corresponding equations are rewritten as non-local Poisson/inhomogeneous Helmholtz equations in compact and non-compact weighted/Riemannian manifolds using static/frequency-domain Green's functions, which can be computed from the heat kernels in the respective manifolds. With the help of the heat kernel estimates, we derive the static Green's function estimates and use them to discuss the regularity. We also present several examples of exact and estimated static/frequency-domain Green's functions.
\end{abstract}

\maketitle

%%%%%%%%%%%%%%%%%%%%%%%%%%%%%%%%%%%%%%%%%%%%%%%%%%%%%%%%%%%%%%%%%%%%%%%%%%%%%%%%
%%%%%%%%%%%%%%%%%%%%%%%%%%%%%%%%%%%%%%%%%%%%%%%%%%%%%%%%%%%%%%%%%%%%%%%%%%%%%%%%
%% Introduction

\section{Introduction}

Non-local exponential operators of the type $e^{-\ell^2\Box}$ appear in physics in various contexts. They arise naturally in effective descriptions of the \textit{string field theory (SFT)} \cite{Witten:1985cc,Freund:1987kt} as well as the \textit{p-adic string theory (PST)} \cite{Brekke:1988dg,Frampton:1988kr}. The non-local modifications of the \textit{general relativity (GR)} achieved through such operators, referred to as the \textit{(ghost-free) infinite-derivative gravity (IDG)}, have also attracted attention \cite{Modesto:2011kw,Biswas:2011ar} (with early works in \cite{Krasnikov:1987yj,Tomboulis:1997gg,Biswas:2005qr}). It has turned out that the infinite-derivative operators such as $e^{-\ell^2\Box}$ have tendency to improve the ultraviolet behavior of GR without introducing ghost degrees of freedom. In particular, they seem to resolve spacetime singularities (based on linearized/weak-field results) and also make the theory (super-)renormalizible.

The presence of an infinite number of derivatives makes the initial value problem very intricate \cite{Barnaby:2007ve,Barnaby:2010kx,Grka2012,Calcagni:2018lyd}. The entire evolution is equivalent to the initial conditions, which are, however, subject to the consistency conditions \cite{Moeller:2002vx}. These conditions manifest themselves in the proposed Hamiltonian formulations usually as (second-class) constraints arising from the field equations \cite{Llosa1994,Gomis:2000gy,Gomis:2003xv,Kolar:2020ezu,Heredia:2021wja}. As a consequence, solving non-local theories can be very intricate because all the standard approaches are not applicable.

The solutions of models inspired by SFT and PST (in the flat space or in the cosmological settings) were studied, for example, in \cite{Brekke:1988dg,Eliezer1989,Sen:2002nu,Moeller:2002vx,Gomis:2003xv,Arefeva:2003mur,Vladimirov:2003kg,Volovich:2003zh,Biswas:2010yx,Barnaby:2010kx,Vladimirov:2006rv,Arefeva:2007wvo,Arefeva:2007xdy}. One method that exploits the exponential form of $e^{-\ell^2\Box}$ relies on recasting the non-local equation into the \textit{heat/diffusion equation} \cite{Calcagni:2007ru,Calcagni:2007ef,Calcagni:2007wy,Calcagni:2008nm} (see also \cite{Calcagni:2018lyd}), which is then often solved iteratively from a trial function using the convolution with the heat kernel or as a boundary value problem \cite{Joukovskaya:2007nq,Joukovskaya:2008zv,Joukovskaya:2008cr} (for an alternative initial value formulation see \cite{Mulryne:2008iq}). If the problem is linearized, then it is common to use the Laplace transforms (on half-line) or/and the Fourier transform (in flat space) or/and some spectral properties of $-\Box$ (in curved space) \cite{Barnaby:2007ve,Barnaby:2008tc,Grka2010,Grka2012,Grka2012v,Heredia:2021pxp} (for a novel Borel transform approach, see \cite{Carlsson2015}). These methods have been also employed for solving IDG in the linearized regime \cite{Biswas:2011ar,Modesto:2010uh,Frolov:2015bta,Frolov:2015usa,Frolov:2015bia,Buoninfante:2018stt,Boos:2018bxf,Kolar:2020bpo,Heredia:2021pxp,frolovemitter2016,Kolar:2021oba}. Known exact solutions are either (generalized) gravitational waves, which lead to  (semi-)linear non-local equations \cite{Kilicarslan:2019njc,Dengiz:2020xbu,Kolar:2021rfl,Kolar:2021uiu} or
bouncing cosmologies with recursive curvature, which lead to non-linear but local equations \cite{Biswas:2005qr,Biswas:2010zk,Biswas:2012bp,Koshelev:2017tvv,Kumar:2020xsl,Kumar:2020xsl}.

Since the non-linear non-local field equations of IDG or SFT/PST (in curved backgrounds) are extremely difficult, it seems very natural to first better understand linear non-local scalar field equations of the form \begin{equation*}
    \boxed{-e^{-\ell^2\square}\square{\phi}=J}
\end{equation*}
in fixed but curved background spacetimes. This paper aims to develop methods for solving such an equation and to provide several examples of i) static scalar fields in static spacetimes and ii) time-dependent scalar fields in ultrastatic spacetimes. We show that the static space/time splitting together with the exponential form of the non-local operator enable us to find Green's functions (and discuss their regularity) via the heat kernels. Furthermore, we try to highlight the importance of mathematical results on the heat kernels and their estimates in compact and non-compact (weighted/Riemannian) manifolds, which are perfectly suited for these non-local problems, but often overlooked by physicists.

This paper is structured as follows: 
In Sec.~\ref{sec:staticspacetimesplitting} we discuss the static space/time splitting, which provides a link between 4-dimensional static spacetimes and 3-dimensional spaces with an additional density (weight). 
In Sec.~\ref{sec:nonlocalscalarfieldtheory}, this viewpoint is utilized for the decomposition of the non-local exponential operator allowing for the representation through the heat kernels on weighted manifolds. In Sec.~\ref{sec:heatkernelsandtheirestimates} we review some exact formulas for the heat kernels and necessary mathematical results on the heat kernel estimates in non-compact weighted manifolds.
In Sec.~\ref{sec:staticgreensfunctions} we solve the non-local equation for static scalar fields in static spacetimes by means of the exact and estimated static Green's functions; we study their regularity and present several explicit examples.
In Sec.~\ref{sec:frequencydomaingreensfunctions} we solve the non-local equation for time-dependent scalar fields in ultrastatic spacetimes using exact frequency-domain Green's functions and we also provide some explicit examples. In Sec.~\ref{sec:conclusions} we conclude the paper with a summary and a discussion of our results.

\subsection*{Remarks on notation}

Let us denote the spaces of sufficiently smooth scalar fields/functions, tensor fields of type $(k,l)$ (with symmetrization marked by parentheses), and densities on the manifold $\mathcal{M}$ by $F\mathcal{M}$, $\bs{T}^k_l \mathcal{M}$, and $\mathfrak{D} \mathcal{M}$, respectively. Boldface is used for tensors while the fraktur font for densities. We use the index-free tensor notation where the dot $\cdot$ indicates the contraction between two adjacent indices and raising/lowering of indices is achieved by means of the \textit{musical isomorphisms} $^{\sharp}$ and $^{\flat}$ \cite{Lee2012}.

We adopt the notation for two-sided estimates from \cite{Grigor:weight}. For positive functions $f$ and $g$ on a set $X$, we write ${f(x) \simeq g(x)}$, ${x\in X}$, if there exists a positive constant $c_{\textrm{m}}$ satisfying ${c_{\textrm{m}}^{-1} \leq {f(x)}/{g(x)} \leq c_{\textrm{m}}}$, ${x\in X}$. Furthermore, we use the notation ${f(x) \asymp g(x, c, \tilde{c})}$, ${x\in X}$ if there are positive constants $c_{\textrm{b}}, \tilde{c}_{\textrm{b}}, c_{\textrm{u}},\tilde{c}_{\textrm{u}}$ for which $g\left(x, c_{\textrm{b}}, \tilde{c}_{\textrm{b}}\right) \leq f(x) \leq g\left(x, c_{\textrm{u}}, \tilde{c}_{\textrm{u}}\right)$, ${x\in X}$. The letter $c$ with various accents and subscripts is reserved for positive constants whose values can change at any occurrence.

%%%%%%%%%%%%%%%%%%%%%%%%%%%%%%%%%%%%%%%%%%%%%%%%%%%%%%%%%%%%%%%%%%%%%%%%%%%%%%%%
%%%%%%%%%%%%%%%%%%%%%%%%%%%%%%%%%%%%%%%%%%%%%%%%%%%%%%%%%%%%%%%%%%%%%%%%%%%%%%%%
%% Static space/time splitting

\section{Static space/time splitting} \label{sec:staticspacetimesplitting}

The starting point of our discussion is the natural space/time splitting that is available in static spacetimes. We introduce a viewpoint in which the 4-dimensional spacetime is regarded as a 3-dimensional space with an additional density.

\subsection{Static spacetimes as weighted 3-manifolds}

Consider a 4-dimensional Lorentzian manifold $(\bar{\mathcal{M}},\bar{\bs{g}})$, ${\bar{\bs{g}}\in \bs{T}_{(2)}^0\bar{\mathcal{M}}}$,\footnote{Bars emphasise spacetime character of the quantities.} which admits a hypersurface-orthogonal timelike Killing vector $\bar{\bs{\xi}}\in\bs{T}_0^1\bar{\mathcal{M}}$, $\lie_{\bar{\bs{\xi}}}\bs{\bar{g}}=0$. Labeling the foliation of the manifold by ${t=\textrm{const.}}$ and assuming ${\bar{\mathcal{M}}=\mathbb{R}\times \mathcal{M}}$, we can write $\bar{\bs{g}}$ as\footnote{We work with the mostly positive metric signature, $(-,+,+,+)$.}
\begin{equation}\label{eq:staticspacetime}
    \bar{\bs{g}}(t,\mathrm{x})=- w^2(\mathrm{x})\bs{\dd} t \bs{\dd} t+ \bs{g}(\mathrm{x})\;, \quad
    t\in\mathbb{R}\;, \; \mathrm{x}\in \mathcal{M}\;.
\end{equation}
Just described spacetimes are called the \textit{static spacetimes}. The notion of \textit{staticity} is with respect to the observers generated by the timelike Killing vector ${\bar{\bs{\xi}}=\bs{\partial}_{t}}$. It is intimately connected with the function $w$ through the norm ${\bar{\bs{\xi}}\cdot \bar{\bs{\xi}}^{\flat}=-w^2}$.

Thanks to the isometry of submanifolds ${t=\textrm{const.}}$, we can capture the full information encoded in $(\bar{\mathcal{M}},\bar{\bs{g}})$ by an arbitrary 3-dimensional Riemannian manifold $(\mathcal{M},\bs{g})$, ${{\bs{g}}\in \bs{T}_{(2)}^0{\mathcal{M}}}$ equipped with an additional positive density ${\mathfrak{w}\in\mathfrak{D}\mathcal{M}}$. Here, $\bs{g}$ is the induced metric on $\mathcal{M}$ and the density $\mathfrak{w}$ is related to the standard metric density ${\mathfrak{g}^{1/2}:=\sqrt{\det g_{ij}}dx^1dx^2dx^3\in\mathfrak{D}\mathcal{M}}$ via the \textit{weight function} ${w}$, 
\begin{equation}
    \mathfrak{w}= w\,\mathfrak{g}^{1/2},
\end{equation}
where ${w\in F\mathcal{M}}$ is positive and sufficiently smooth. The triplet ${(\mathcal{M},\bs{g},\mathfrak{w})}$ is often referred to as the \textit{weighted manifold}. 

A special subclass of spacetimes corresponding to the unweighted case ${w=1}$ arises if the Killing vector $\bar{\bs{\xi}}$ is also covariantly constant, ${\bar{\bs{\nabla}}\bar{\bs{\xi}}=0}$. Then the spacetime metric $\bar{\bs{g}}$ reduces to
\begin{equation}\label{eq:ultrastaticspacetime}
    \bar{\bs{g}}(t,\mathrm{x})=-\bs{\dd} t \bs{\dd} t+ \bs{g}(\mathrm{x})\;,
\end{equation}
which is commonly referred to as the \textit{ultrastatic spacetime} \cite{Stephani2003} (see also \cite{Sonego:2010vy} and references therein). The ultrastaticity basically means that observers in all ${\mathrm{x}\in\mathcal{M}}$ associated with $\bar{\bs{\xi}}$ share a common proper time $t$ (in contrast to other static spacetimes).

Let us introduce some geometric quantities that we can calculate in ${(\mathcal{M},\bs{g},\mathfrak{w})}$. In what follows, $(\mathcal{M},\bs{g})$ is always assumed to be connected and geodesically complete (without boundary). This allows us to compute the \textit{(shortest) geodesic distance} $D_{\mathrm{xy}}$ between every two points ${\mathrm{x},\mathrm{y}\in\mathcal{M}}$. Consequently, we can define a \textit{geodesic ball} ${\mathcal{B}(\mathrm{x},\rho)\subset\mathcal{M}}$ of radius $\rho$ centered at ${\mathrm{x}\in\mathcal{M}}$. Its \textit{volume} $V(\mathrm{x}, \rho)$ and \textit{surface} $S(\mathrm{x}, \rho)$ are
\begin{equation}
    V(\mathrm{x}, \rho):=\int\limits_{\mathclap{\mathcal{B}(\mathrm{x},\rho)}}\mathfrak{g}^{1/2}\;,
    \quad 
    S(\mathrm{x}, \rho):=\int\limits_{\mathclap{\partial\mathcal{B}(\mathrm{x},\rho)}}\mathfrak{g}^{1/2}|_{\partial\mathcal{B}(\mathrm{x},\rho)}\;,
\end{equation}
which are measured with the metric density $\mathfrak{g}^{1/2}$. Since we have access to the density $\mathfrak{w}$, it is very convenient to introduce also the \textit{weighted volume} $V_{\mathfrak{w}}(\mathrm{x}, \rho)$ and \textit{weighted surface} $S_{\mathfrak{w}}(\mathrm{x}, \rho)$ of ${\mathcal{B}(\mathrm{x},\rho)}$,
\begin{equation}
    V_{\mathfrak{w}}(\mathrm{x}, \rho):=\int\limits_{\mathclap{\mathcal{B}(\mathrm{x},\rho)}}\mathfrak{w}\;,
    \quad 
    S_{\mathfrak{w}}(\mathrm{x}, \rho):=\int\limits_{\mathclap{\partial\mathcal{B}(\mathrm{x},\rho)}}\mathfrak{w}|_{\partial\mathcal{B}(\mathrm{x},\rho)}\;.
\end{equation}

\subsection{Wave operator}

In our $3$-dimensional description, an arbitrary scalar field ${\bar{\phi}\in F \bar{\mathcal{M}}}$ can be realized as a set of scalar fields ${\phi^t\in F\mathcal{M}}$ parametrized by ${t\in\mathbb{R}}$, ${\phi^t(\mathrm{x})=\bar{\phi}(t,\mathrm{x})}$. The action of the \textit{wave operator} ${\bar\square:=\bar{\bs\nabla}\cdot\bar{\bs\dd}^{\sharp}}$ on $\bar{\phi}$ can then be understood as
\begin{equation}\label{eq:boxdecomp}
    \bar\square\bar{\phi}(t,\mathrm{x}) =\big({-}w^{-2}\partial_{t}^2+\triangle_{\mathfrak{w}}\big)\phi^t(\mathrm{x})\;,
\end{equation}
where we introduced the \textit{weighted Laplace operator}  \cite{chaveletal:1991,Davies1992},
\begin{equation}
\begin{aligned}
    \triangle_{\mathfrak{w}} &:=\mathfrak{w}^{-1}{\di}\big(\mathfrak{w}\,{\bs\dd}^{\sharp}\big)=w^{-1}{\bs\nabla}\cdot\big(w\,{\bs\dd}^{\sharp}\big)
    \\
    &=\triangle+\big(\bs{\dd}\log w\big)\cdot \bs{\dd}^{\sharp}\;,
\end{aligned}
\end{equation}
with ${\triangle:={\bs\nabla}\cdot{\bs\dd}^{\sharp}}$ denoting the standard (unweighted) Laplace operator. If the spacetime is ultrastatic, i.e., ${w=1}$, then ${\triangle_{\mathfrak{w}}=\triangle}$ and the wave operator reduces to ${\bar\square =-\partial_{t}^2+\triangle}$. 

Another advantage of (ultra)static spacetimes is that we can always go from the time domain to the frequency domain ${t\to\omega}$ (and back) by means of the standard Fourier transform,
\begin{equation}\label{eq:FT}
    \hat{f}^{\omega}=\frac{1}{\sqrt{2\pi}}\int\limits_{\mathbb{R}}\!d t \, e^{-i\omega t}{f}^{t}\;, \quad     f^t=\frac{1}{\sqrt{2\pi}}\int\limits_{\mathbb{R}}\!d\omega \, e^{i\omega t}\hat{f}^{\omega}\;,
\end{equation}
which replaces ${-\partial_t^2\to\omega^2}$ and ${\phi^t\to\hat{\phi}^{\omega}}$ on the right-hand side of \eqref{eq:boxdecomp}. 

If $\bar{\phi}$ is independent of $t$, i.e., ${\lie_{\bar{\bs{\xi}}}\bar{\phi}=0}$, then we call it the \textit{static scalar field} and describe it by a single scalar field ${\phi\in F\mathcal{M}}$. When acting on a static scalar field, the wave operator is given simply by the weighted Laplace operator ${\bar\square=\triangle_{\mathfrak{w}}}$.

%%%%%%%%%%%%%%%%%%%%%%%%%%%%%%%%%%%%%%%%%%%%%%%%%%%%%%%%%%%%%%%%%%%%%%%%%%%%%%%%
%%%%%%%%%%%%%%%%%%%%%%%%%%%%%%%%%%%%%%%%%%%%%%%%%%%%%%%%%%%%%%%%%%%%%%%%%%%%%%%%
%% Non-local scalar field theory

\section{Non-local scalar field theory} \label{sec:nonlocalscalarfieldtheory}

The space/time splitting of the wave operator $\bar{\Box}$ (and the Fourier transform in time) is especially useful in the study of non-local scalar fields. It allows us to rewrite the field equations in terms of operators $e^{\tau\triangle_{\mathfrak{w}}}$, which can then be represented through the heat kernels on weighted manifolds.

\subsection{Field equation}

Consider a non-local theory of a scalar field ${\bar\phi\in F\bar{\mathcal{M}}}$ in a fixed spacetime $(\bar{\mathcal{M}},\bar{\bs{g}})$ sourced by ${\bar{J}\in F\bar{\mathcal{M}}}$,
\begin{equation}
    \mathcal{S}[\bar{\phi}]=\int\limits_{\bar{\mathcal{M}}}\bar{\mathfrak{g}}^{1/2}\left(\frac12 \bar{\phi}e^{-\ell^2\bar\square}\bar\square\bar{\phi}+\bar{J}\bar{\phi}\right)\;,
\end{equation}
where the non-local exponential operator can be understood either via the infinite sum of derivatives or via the spectral resolution $\bar{E}_{\bar\lambda}$ of $-\bar{\square}$,
\begin{equation}\label{eq:expinfsum}
    e^{\tau\bar\square}:=
        \sum_{k=0}^{\infty}\frac{\tau^k}{k!}\bar{\square}^k:=\int\limits_{\mathbb{R}}  \!d \bar{E}_{\bar\lambda}\,e^{-\tau\bar\lambda}\;, \quad \tau\in\mathbb{R}\;,
\end{equation}
provided that $(\bar{\mathcal{M}},\bar{\bs{g}})$ admits such representations for certain spaces of functions. The parameter ${\ell>0}$ is called the \textit{(length) scale of non-locality}. The local theory is recovered in the limit ${\ell\to 0}$. By taking the functional derivative of $\mathcal{S}$ with respect to $\bar{\phi}$, we may derive the field equation, which is of the of the non-local inhomogeneous wave equation,
\begin{equation}\label{eq:nlequation}
    -e^{-\ell^2\bar\square}\bar{\square}\bar{\phi}=\bar{J}\;.
\end{equation}

Considering a static spacetime, we can rewrite this equation in the spirit of the previous section as
\begin{equation}\label{eq:nonlocaleq}
    e^{\ell^2\left(w^{-2}\partial_t^2-\triangle_{\mathfrak{w}}\right)}\big(w^{-2}\partial_t^2-\triangle_{\mathfrak{w}}\big)\phi^t=J^t\;.
\end{equation}
In this paper, we will study two special subcases, in which the operators in the argument of the exponential commutes, ${\big[w^{-2}\partial_t^2,\triangle_{\mathfrak{w}}\big]\phi^t=0}$:
\begin{enumerate}[label=\roman*)]
\item If the scalar field and the spacetime are both static, then \eqref{eq:nonlocaleq} reduces to the \textit{non-local Poisson's equation}:
\begin{equation}\label{eq:redeq1}\boxed{
    -\triangle_{\mathfrak{w}}\phi=e^{\ell^2\triangle_{\mathfrak{w}}}J\;.}
\end{equation}
\item If the scalar field is time-dependent but the spacetime is ultrastatic, then we can recast the equation \eqref{eq:nonlocaleq} to the form of the \textit{non-local inhomogeneous Helmholtz equation} (in frequency domain):
\begin{equation}\label{eq:redeq2}
    \boxed{-\big(\triangle+\omega^2\big)\hat{\phi}^{\omega}=e^{\ell^2\omega^2}e^{\ell^2\triangle}\hat{J}^{\omega}\;.}
\end{equation}
\end{enumerate}
Notice that we assumed that the non-local exponential operator can be inverted without affecting the solutions of the field equation \cite{Barnaby:2007ve,Barnaby:2008tc}. Because of this the non-local operator with the (weighted) Laplace operator appears on the right-hand side with the positive sign in the exponent. As a consequence, its action can be formulated by means of the heat kernel (below).

It is vital to stress that the non-local exponential operators defined through infinite sums of derivatives, spectral representations, and the heat kernels may differ on certain (non-analytic) functions \cite{Moeller:2002vx}. Here, we assume that we work in a subspace of functions where all three definitions are equivalent (and the exponential operator is invertible). Explicit identification of such spaces is rather difficult and goes beyond the scope of this paper (see \cite{Heredia:2021pxp}, for the discussion of this problem in the Minkowski spacetime). We will be quite sloppy in this regard and just write $F\bar{\mathcal{M}}$ or $F\mathcal{M}$ for such optimal spaces of functions.

\subsection{Non-local operator via heat kernel}

In curved spaces, it is very efficient to characterize the action of $e^{\tau\triangle_{\mathfrak{w}}}$, ${\tau\geq0}$, by means of the heat kernels. Consider the \textit{heat/diffusion equation} on ${[0,\infty)\times \mathcal{M}}$ for an arbitrary initial condition ${\Psi_0\in F\mathcal{M}}$,
\begin{equation}\label{eq:heateq}
    \triangle_{\mathfrak{w}} \Psi_{\tau}(\mathrm{x})=\partial_{\tau} \Psi_{\tau}(\mathrm{x})\;, 
\end{equation}
where ${\tau\ge 0}$ is an evolution parameter. The solution is often expressed using the \textit{heat kernel} ${K_{\tau}\in F(\mathcal{M}\times\mathcal{M})}$ \cite{grigoryanheat},
\begin{equation}\label{eq:PhitjeHK}
    \Psi_{\tau}(\mathrm{x}) =:\langle K_{\tau}(\mathrm{x},\cdot),\Psi_0\rangle_{\mathfrak{w}}\;,
\end{equation}
where we introduced the weighted inner product of two scalar fields ${f_1,f_2\in F\mathcal{M}}$,
\begin{equation}\label{eq:innerproduct}
    \langle f_1,f_2\rangle_{\mathfrak{w}} :=\int\limits_{\mathclap{\mathrm{x}\in \mathcal{M}}}\mathfrak{w}(\mathrm{x})f_1(\mathrm{x})f_2(\mathrm{x})\;.
\end{equation}
Let us point out that the heat kernel is always symmetric, ${K_{\tau}(\mathrm{x},\mathrm{y})=K_{\tau}(\mathrm{y},\mathrm{x})}$, but we will not symmetrize it in many cases for brevity reasons.

Denoting the spectral resolution of the weighted Laplace operator ${-\triangle_{\mathfrak{w}}}$ by ${E_{\lambda}}$, we may write
\begin{equation}\label{eq:specresofexp}
    e^{\tau\triangle_{\mathfrak{w}}}:=\int\limits_{0}^{\infty}\! d E_{\lambda}\; e^{-\tau\lambda}\;.
\end{equation}
Therefore the action of the non-local exponential operator $e^{\tau\triangle_{\mathfrak{w}}}$ on an arbitrary scalar field ${\Psi_0\in F\mathcal{M}}$ can be equivalently characterized by means of the solution $\Psi_{\tau}$, i.e., diffused with parameter $\tau$,
\begin{equation}\label{eq:defexpnl}
    e^{\tau\triangle_{\mathfrak{w}}}\Psi_0=\Psi_{\tau}\;, 
\end{equation}
because the differentiation with respect to the parameter~$\tau$, ${\partial_{\tau}e^{-\tau\lambda}=-\lambda e^{-\tau\lambda}}$, exactly reproduces \eqref{eq:heateq}. Comparing \eqref{eq:defexpnl} with \eqref{eq:PhitjeHK} we also obtain the distributional formula for the heat kernel
\begin{equation}\label{eq:Kjeenadelta}
    K_{\tau}(\cdot,\mathrm{y})=e^{\tau\triangle_{\mathfrak{w}}}\delta_{\mathrm{y}}\;,
\end{equation}
which also implies that $K_{\tau}$ satisfies the heat equation in both variables. Here, $\delta_{\mathrm{y}}$ is the Dirac delta distribution defined by
\begin{equation}
    \langle \delta_{\mathrm{y}},f\rangle_{\mathfrak{w}}:=f(\mathrm{y})\;,
\end{equation}
where $\langle T,f\rangle_{\mathfrak{w}}$ is now understood as an action of the distribution (linear functional) $T$ on the test function $f$. This bracket reduces back to the inner product \eqref{eq:innerproduct} for distributions associated with ordinary functions. The set of operators $e^{\tau\triangle_{\mathfrak{w}}}$, ${\tau\geq0}$, is often referred to as the \textit{heat semi-group} due to the identity
\begin{equation}\label{eq:semigroup}
    e^{\tau_1\triangle_{\mathfrak{w}}}e^{\tau_2\triangle_{\mathfrak{w}}}=e^{(\tau_1+\tau_2)\triangle_{\mathfrak{w}}}\;,
\end{equation}
which translates to the following identity for the heat kernel
\begin{equation}\label{eq:semigroup2}
    K_{\tau_1+\tau_2}(\mathrm{x},\mathrm{z})=\langle K_{\tau_1}(\mathrm{x},\cdot)K_{\tau_2}(\cdot,\mathrm{z})\rangle_{\mathfrak{w}}\;.
\end{equation}

Let us also mention two important limits of the heat kernel \cite{Grigor:weight},
\begin{equation}\label{eq:limitsHK}
\begin{aligned}
    \lim_{\tau\to 0}\tau\log K_{\tau}(\mathrm{x},\mathrm{y}) &=-\tfrac{1}{4} D_{\mathrm{xy}}^2\;, \\
    \lim_{\tau\to \infty}\frac{1}{\tau}\log K_{\tau}(\mathrm{x},\mathrm{y}) &=-\lambda_{\textrm{min}}\;,
\end{aligned}
\end{equation}
where $\lambda_{\textrm{min}}$ denotes the bottom of the spectrum of ${-\triangle_{\mathfrak{w}}}$. The former can be made more explicit if $D_{\mathrm{xy}}$ is sufficiently small. Then $K_{\tau}$ always approaches the heat kernel of the Euclidean 3-space~$(\mathbb{R}^3,\bs{g}_{\textrm{euc}})$ (given by \eqref{eq:eucHK} below) multiplied by a positive function ${f\in F(\mathcal{M}\times\mathcal{M})}$,
\begin{equation}\label{eq:HKtau0approx}
    K_{\tau}(\mathrm{x},\mathrm{y}) \sim f(\mathrm{x},\mathrm{y})K_{\tau}^{\textrm{euc}}(\mathrm{x},\mathrm{y})\;, \quad \tau\to 0\;.
\end{equation}

If the manifold is compact, then $-\triangle_{\mathfrak{w}}$ has pure (non-negative) point spectrum ${\lambda_k}$, ${k\in\mathbb{N}_0}$ (counted with multiplicity). Moreover, $\lambda_k$ can be sorted so that the sequence grows and ${\lambda_k\to\infty}$ for ${k\to\infty}$. The corresponding eigenfunctions~${\psi_k\in F\mathcal{M}}$,
\begin{equation}
    -\triangle_{\mathfrak{w}}\psi_k=\lambda_k\psi_k\;,
\end{equation}
form a complete orthonormal set on $F\mathcal{M}$ with the inner product \eqref{eq:innerproduct}, \begin{equation}\label{eq:normalization}
    \langle\psi_j,\psi_k\rangle=\delta_{jk}\;,
\end{equation}
meaning that we can expand an arbitrary function ${f\in F\mathcal{M}}$ as
\begin{equation}\label{eq:expansionoff}
    f(\mathrm{x})=\sum_{k=0}^{\infty} f_{k}\psi_{k}(\mathrm{x})\;,
    \quad
    f_{k}:=\langle f,\psi_{k}\rangle_{\mathfrak{w}}\;.
\end{equation}
Furthermore, the minimum principle implies that the constant function is the only harmonic function on compact manifold. Due to the normalization \eqref{eq:normalization} we find 
\begin{equation}\label{eq:firsteigen}
    \lambda_0=0\;, \quad \psi_0=\frac{1}{\sqrt{V_{\mathfrak{w}}^{\mathcal{M}}}}\;,
\end{equation}
where $V_{\mathfrak{w}}^{\mathcal{M}}:=\int_{\mathcal{M}}\mathfrak{w}$ denotes the weighted volume of $\mathcal{M}$, which is finite because $w$ and $\bs{g}$ are bounded on $\mathcal{M}$ (thanks to continuity). The action of the non-local exponential operator \eqref{eq:specresofexp} reduces to the infinite sum
\begin{equation}\label{eq:infsumeig}
    e^{\tau\triangle_{\mathfrak{w}}}f=\sum_{k=0}^{\infty}e^{-\tau\lambda_k} f_k\psi_k\;,
\end{equation}
which can be easily translated to a direct relation for the heat kernel if we expand $f$ in ${e^{\tau\triangle_{\mathfrak{w}}}f(\mathrm{x})=\langle K_{\tau}(\mathrm{x},\cdot),f\rangle_{\mathfrak{w}}}$ by means of \eqref{eq:expansionoff} with \eqref{eq:normalization},
\begin{equation}\label{eq:heatkerneleigen}
    K_{\tau}(\mathrm{x},\mathrm{y})=\sum_{k=0}^{\infty}e^{-\tau\lambda_k}\psi_{k}(\mathrm{x})\psi_{k}(\mathrm{y})\;.
\end{equation}
The sum converges absolutely and uniformly for ${\tau\geq\varepsilon}$ by Weierstrass $m$-test with the convergent majorant series ${\sum_{k=0}^{\infty}m_k(\varepsilon)}$ for any ${\varepsilon>0}$ \cite{grigoryanheat}.

%%%%%%%%%%%%%%%%%%%%%%%%%%%%%%%%%%%%%%%%%%%%%%%%%%%%%%%%%%%%%%%%%%%%%%%%%%%%%%%%
%%%%%%%%%%%%%%%%%%%%%%%%%%%%%%%%%%%%%%%%%%%%%%%%%%%%%%%%%%%%%%%%%%%%%%%%%%%%%%%%
%% Non-local scalar field theory

\section{Heat kernels and their estimates} \label{sec:heatkernelsandtheirestimates}

Before we proceed to solving equations \eqref{eq:redeq1} and \eqref{eq:redeq2}, we need to review some exact formulas for heat kernels together with important mathematical results on heat kernel estimates in non-compact weighted manifolds. For more details, we refer the reader to the reviews \cite{Grigor:weight,Grigoryan1999,SaloffCoste} and the textbook \cite{grigoryanheat}.

\subsection{Exact expressions}

Let us begin with some important 3-dimensional (unweighted) Riemannian manifolds for which the heat kernels are known exactly. Such examples involve mainly the maximally symmetric spaces (with 6 Killing vectors), i.e., the space of constant curvature. 

Of course, the simplest and best-known example is the Euclidean 3-space~$(\mathbb{R}^3,\bs{g}_{\textrm{euc}})$ (vanishing curvature),\footnote{Spherically symmetric metrics are written in spherical coordinates for later convenience. All expressions for the heat kernels are coordinate independent.}
\begin{equation}\label{eq:geuc}
    \bs{g}_{\textrm{euc}}:=\bs{\dd} \rho \bs{\dd} \rho+\rho^2(\bs{\dd} \vartheta \bs{\dd} \vartheta+\sin^2\vartheta\,\bs{\dd} \varphi \bs{\dd} \varphi)\;.
\end{equation}
Here, the Laplace operator $-\triangle$ has a continuous non-negative spectrum, ${[0,\infty)}$. The heat kernel has the form of the Gaussian function with the variance $\tau$,
\begin{equation}\label{eq:eucHK}
    K_{\tau}^{\textrm{euc}}(\mathrm{x},\mathrm{y}) = \frac{1}{(4 \pi \tau)^{\frac{3}{2}}} \exp\left(-\frac{D_{\mathrm{xy}}^2}{4 \tau}\right)\;,
\end{equation}
where $D_{\mathrm{xy}}$ is now the standard Euclidean distance. 

Another non-compact Riemannian manifold with an exact formula is the hyperbolic 3-space $(\mathbb{R}^3,\bs{g}_{\textrm{hyp}})$ (negative constant curvature),
\begin{equation}\label{eq:ghyp}
    \bs{g}_{\textrm{hyp}}:=\bs{\dd} \rho \bs{\dd} \rho+A^2\sinh^2\left(\tfrac{\rho}{A}\right)(\bs{\dd} \vartheta \bs{\dd} \vartheta+\sin^2\vartheta\,\bs{\dd} \varphi \bs{\dd} \varphi)\;,
\end{equation}
where $A>0$ is an arbitrary constant characterizing the value of the curvature. The heat kernel is given by \cite{Mckean1970AnUB,DaviesMandouvalos:1988,GrigoryanMasakazu:1998}
\begin{equation}\label{eq:Khyp}
    K_{\tau}^{\textrm{hyp}}(\mathrm{x},\mathrm{y}) =\frac{1}{(4 \pi \tau)^{\frac32}}\frac{\frac{D_{\mathrm{xy}}}{A}}{\sinh \left(\frac{D_{\mathrm{xy}}}{A}\right)}  \exp\left(-\frac{D_{\mathrm{xy}}^2}{4 \tau}-\frac{\tau}{A^2}\right)\;.
\end{equation}
The additional term $-\tau/A^2$ arises thanks to the fact that the spectrum starts above zero, ${[1/A^2,\infty)}$, with  ${\lambda_{\textrm{min}}=1/A^2}$, see the second limit of \eqref{eq:limitsHK}. 

Moving on to the compact Riemannian manifolds, the exact expression is known for the 3-sphere~$(\mathbb{S}^3,\bs{g}_{\textrm{sph}})$ (positive constant curvature),
\begin{equation}\label{eq:gsph}
    \bs{g}_{\textrm{sph}}:=\bs{\dd} \rho \bs{\dd} \rho+B^2\sin^2\left(\tfrac{\rho}{B}\right)(\bs{\dd} \vartheta \bs{\dd} \vartheta+\sin^2\vartheta\,\bs{\dd} \varphi \bs{\dd} \varphi)\;,
\end{equation}
where $B>0$ denotes the radius of the 3-sphere. Its heat kernel reads \cite{David:2009xg,Camporesi:1990wm}
\begin{equation}\label{eq:Ksph}
    K_{\tau}^{\textrm{sph}}(\mathrm{x},\mathrm{y})=\sum_{k=0}^{\infty}\tfrac{(k+1) \,\sin \left((k+1) \frac{D_{\mathrm{xy}}}{B}\right)}{2 \pi^{2}B^3\sin\left( \frac{D_{\mathrm{xy}}}{B}\right)} e^{-k(k+2) \frac{\tau}{B^2}}\;.
\end{equation}
This sum converges absolutely and uniformly for ${\tau\geq\varepsilon}$ for any ${\varepsilon>0}$ as we anticipated. Since the spectrum of he Laplace operator $-\triangle$ is discrete and consisting of the eigenvalues ${\lambda_k=k(k+2)/B^2}$, ${k\in\mathbb{N}_0}$ with the eigenfunctions being 3-dimensional spherical harmonics $\psi_{k,j}$. The formula \eqref{eq:Ksph} is obtained by means of \eqref{eq:heatkerneleigen} with the use of the \textit{addition theorem} \cite{Wen1985},\footnote{This identity typically appears with the Gegenbauer polynomial on the right-hand side, ${C_k^{(1)}(\cos x)=\frac{\sin ((k+1) x)}{\sin x}}$.}
\begin{equation}\label{eq:addtheo}
    \psi_{k}(\mathrm{x})\psi_{k}(\mathrm{y})=\sum_{j=1}^{\mathclap{(k+1)^2}}\psi_{k,j}(\mathrm{x})\psi_{k,j}(\mathrm{y})=\tfrac{(k+1)\sin \left((k+1) \frac{D_{\mathrm{xy}}}{B}\right)}{2\pi^2B^3\sin \left(\frac{D_{\mathrm{xy}}}{B}\right)}\;.
\end{equation}

Note that the expressions $K_{\tau}^{\textrm{hyp}}$ and $K_{\tau}^{\textrm{sph}}$ approach $K_{\tau}^{\textrm{euc}}$ for large values of $A$ and $B$ (the latter can be verified numerically). Unfortunately, the heat kernels for more complicated (less symetrical) Riemannian/weighted manifolds are not known in general. Luckily, it is often sufficient to work just with the global estimates (upper and/or lower bounds) instead, which have been studied extensively in the mathematical literature.

\subsection{Li--Yau estimate}

It turns out that on a large number of non-compact weighted manifolds $(\mathcal{M},\bs{g},\mathfrak{w})$, the heat kernel can be globally estimated from both sides by (see \cite{Grigor:weight} and references therein)
\begin{equation}\label{eq:LiYaugen}
    K_{\tau}(\mathrm{x},\mathrm{y}) \asymp \frac{c}{V_{\mathfrak{w}}(\mathrm{x}, \sqrt{\tau})} \exp \left(-\tilde{c} \frac{D_{\mathrm{xy}}^2}{4\tau}\right)\;.
\end{equation}
This formula is known as the \textit{(weighted) Li--Yau estimate}. (The constant $\tilde{c}_2$ in the upper bound can be taken arbitrarily close to $1$.) It was first proven for Riemannian manifolds $(\mathcal{M},\bs{g})$ (compact or non-compact) with the \textit{non-negative Ricci curvature}, ${\bs{v}\cdot\bs{Ric}\cdot\bs{v}\geq 0}$, ${\forall \bs{v}\in\bs{T}_0^1\mathcal{M}}$ in the seminal work \cite{Li1986}, where ${\bs{Ric}\in\bs{T}_{(2)}^0\mathcal{M}}$ is the Ricci tensor. Following \cite{Li1986}, this result was extended to many other cases including weighted manifolds $(\mathcal{M},\bs{g},\mathfrak{w})$. 

Such a generalization makes use of two alternative characterizations of weighted manifolds satisfying \eqref{eq:LiYaugen}. The weighted Li--Yau estimate is, on one hand, fully equivalent to the \textit{Harnack inequality}\footnote{${\forall\mathcal{B}(\mathrm{x},\rho)}$, ${\forall\Psi_{\tau}>0}$ solving the heat equation in the cylinder ${\mathcal{C}:=(0,\rho^2)\times\mathcal{B}(\mathrm{x},\rho)}$,
\begin{equation*}
    \sup_{\mathrm{y}\in\mathcal{C}_{-}} \Psi_{\tau}(\mathrm{y}) \leq c \inf_{\mathrm{y}\in\mathcal{C}_{+}} \Psi_{\tau}(\mathrm{y})\;,
\end{equation*}
where ${\mathcal{C}_{-}:=\big(\tfrac{\rho^2}{4},\tfrac{\rho^2}{2}\big)\times\mathcal{B}\big(\mathrm{x},\tfrac{\rho}{2}\big)}$ and ${\mathcal{C}_{+}:=\big(\tfrac{3\rho^2}{4},\rho^2\big)\times\mathcal{B}\big(\mathrm{x},\tfrac{\rho}{2}\big)}$.
} and, on the other hand, to the \textit{Poincar\'e inequality}\footnote{${\exists\varepsilon\in(0,1)}$: $\forall\mathcal{B}(\mathrm{x},\rho)$ and ${\forall f\in F\mathcal{B}(\mathrm{x},\rho)}$,
\begin{equation*}
    \inf_{s\in\mathbb{R}}\int_{{\mathcal{B}(\mathrm{x},\varepsilon\rho)}}\!\!\!\!\!\!\!\!\!\!\!\!\!\!\!\mathfrak{w}\,\big(f-s\big)^2\leq c \rho^2\int_{{\mathcal{B}(\mathrm{x},\rho)}}\!\!\!\!\!\!\!\!\!\!\!\!\mathfrak{w}\,\bs{\dd}f\cdot\bs{\dd}^{\sharp}f\;.
\end{equation*}.} together with the \textit{volume doubling} property,
\begin{equation}\label{eq:voldoub}
    V_{\mathfrak{w}}(\mathrm{x},2\rho)\leq c V_{\mathfrak{w}}(\mathrm{x},\rho)\;.
\end{equation}
The last condition has an important consequence called the \textit{reverse volume doubling},
\begin{equation}\label{eq:voldoubcons}
    \frac{V_{\mathfrak{w}}(\mathrm{x},\rho_2) }{V_{\mathfrak{w}}(\mathrm{x},\rho_1)}\leq c\left(\frac{\rho_2}{\rho_1}\right)^{\kappa}\;, \quad \rho_2>\rho_1>0\;,
\end{equation}
for some ${\kappa>0}$. Assuming that \eqref{eq:LiYaugen} holds for a non-compact Riemannian manifold, one can show (based on the equivalent characterizations above) that it holds also for its weighted counterpart whenever the weight function $w$ is bounded,
\begin{equation}\label{eq:wsim1}
    w(\mathrm{x})\simeq1\;, \quad  \forall x\in\mathcal{M}\;.
\end{equation}

In fact, this result can be further generalized to unbounded $w$ if we introduce another geometric notion. Let us fix a reference point ${\mathrm{o}\in \mathcal{M}}$, called the \textit{origin}, together with the notation
\begin{equation}
\begin{aligned}
    |\mathrm{x}|:=D_{\mathrm{xo}}\;,  
    \quad
    & & V(\rho) &:=V(\mathrm{o},\rho)\;, & V_{\mathfrak{w}}(\rho) &:=V_{\mathfrak{w}}(\mathrm{o},\rho)\;,
    \\ 
    & & S(\rho) &:=S(\mathrm{o},\rho)\;, & S_{\mathfrak{w}}(\rho) &:=S_{\mathfrak{w}}(\mathrm{o},\rho)\;.
\end{aligned}
\end{equation}
We say that $(\mathcal{M},\bs{g})$ has \textit{relatively connected annuli} if there exists a constant ${c>1}$ such that for any two points ${\mathrm{x},\mathrm{y}\in \mathcal{M}}$ with large enough ${|\mathrm{x}|=|\mathrm{y}|=\rho}$ there exists a continuous path from $\mathrm{x}$ to $\mathrm{y}$ that is fully contained within the annulus $\mathcal{B}(\mathrm{o},c\rho)\setminus  \mathcal{B}(\mathrm{o},c^{-1}\rho)$. The statement that relates the weighted and unweighted Li--Yau estimates then remains true on any manifold with relatively connected annuli if we replace the condition \eqref{eq:wsim1} by more general conditions \cite{GrigSaloff,Grigor:weight}
\begin{equation}\label{eq:weightcond}
\begin{gathered}
    w(\mathrm{x}) \simeq \breve{w}(|\mathrm{x}|)\;,
    \quad \forall\mathrm{x}\in\mathcal{M}\;,
    \\
    \int\limits_c^{\rho} \frac{d\tilde{\rho}}{\tilde{\rho}}\breve{w}(\tilde{\rho}) V(\tilde{\rho}) \simeq \breve{w}(\rho) V(\rho)\;, 
    \quad  \forall\rho>2c>0\;,
\end{gathered}
\end{equation}
where we denoted ${\breve{w}(\rho):=\sup_{|\mathrm{x}|=\rho}w(\mathrm{x})}$. Moreover, conditions \eqref{eq:weightcond} always imply the following relation between the weighted and unweighted volumes
\begin{equation}\label{eq:volrel}
    V_{\mathfrak{w}}(\mathrm{x}, \rho)\simeq \breve{w}(|\mathrm{x}|+\rho)\,V(\mathrm{x}, \rho)\;.
\end{equation}

If $V(\rho)\simeq \rho^{\alpha}$ and $\breve{w}(\rho)\simeq \rho^{\beta}$ for large $\rho$, then \eqref{eq:weightcond} is met for ${\alpha+\beta>0}$. Note that the unweighted volume growth is already constrained to $\alpha\in[1,3]$ for non-compact Riemannian manifolds ${(\mathcal{M},\bs{g})}$ of non-negative Ricci curvature according to the \textit{Calabi--Yau} and \textit{Bishop--Gromov bounds} (see e.g. \cite{Leandro2019}),\footnote{The upper bound ${V(\mathrm{x},\rho)=c_2\rho^3}$ corresponds to the Euclidean-like volume growth, which can be achieved, for example, by any conformally deformed Euclidean geometries ${\bs{g}=\Omega^2\bs{g}_{\textrm{euc}}}$, with ${V^{\mathcal{M}}=\infty}$ and ${\int_{\mathcal{M}} \mathfrak{g}^{1/2} R<\infty}$ \cite{Carron2020}.}
\begin{equation}
    c_1\rho\leq V(\mathrm{x},\rho)\leq c_2\rho^3\;,
\end{equation}
for sufficiently large $\rho$.

\subsection{Spherical symmetry}

The problem simplifies considerably if we focus on certain \textit{spherically symmetric weighted spaces} (with 3 Killing vectors). We define them as the weighted manifolds $(\mathbb{R}^3,\bs{g}_{\mathrm{sym}},\mathfrak{s})$, where
\begin{equation}
\begin{aligned}
    \bs{g}_{\textrm{sym}} &:=\bs{\dd} \rho \bs{\dd} \rho+\varrho^2(\rho)(\bs{\dd} \vartheta \bs{\dd} \vartheta+\sin^2\vartheta\,\bs{\dd} \varphi \bs{\dd} \varphi)\;, 
    \\
    \mathfrak{s} &:=s\,\mathfrak{g}_{\textrm{sym}}^{1/2}=s(\rho)\varrho^2(\rho)\sin\vartheta \;d\rho d\vartheta d\varphi\;,
\end{aligned}
\end{equation}
with $\varrho$ and $s$ being two completely arbitrary positive functions satisfying
\begin{equation}\label{eq:condpocatek}
    \varrho(\rho)=\rho+O(\rho^3)\;, \quad  s(\rho)=1+O(\rho^2)\;, \quad \rho\to0\;.
\end{equation}

Just described weighted manifolds are non-compact and have relatively connected annuli with the origin ${\mathrm{o}\in \mathbb{R}^3}$ set at ${\rho=0}$. The additional conditions \eqref{eq:condpocatek} guarantee that the metric is well behaved at $\mathrm{o}$; in particular, $(\mathbb{R}^3,\bs{g}_{\mathrm{sym}},\mathfrak{s})$ becomes $(\mathbb{R}^3,\bs{g}_{\textrm{euc}})$ in the vicinity of the origin.\footnote{From the 4-dimensional viewpoint, these spaces correspond to the spherically symmetric static spacetimes (with 4 Killing vectors) with spatial sections diffeomorphic to ${\mathbb{R}^3}$, which approach Minkowski spacetime near the origin.} The coordinate $\rho$ has the meaning of the geodesic distance in the radial direction $\bs{\partial}_{\rho}$ from the origin, ${\rho=|\mathrm{x}|}$. The weighted volume and surface of a geodesic ball $\mathcal{B}(\mathrm{o},\rho)$ are given by
\begin{equation}
    V_{\mathfrak{s}}(\rho)=\int\limits_{0}^{\rho}\! d\tilde{\rho}\, S_{\mathfrak{s}}(\tilde{\rho})\;, \quad S_{\mathfrak{s}}(\rho)=V'_{\mathfrak{s}}(\rho)=4\pi\, s(\rho)\varrho^2(\rho)\;.
\end{equation}
Since the Weyl tensor vanishes identically in 3 dimensions, the Riemannian curvature is described purely by the Ricci tensor (and Ricci scalar),
\begin{equation}
\begin{aligned}
    \bs{Ric}_{\textrm{sym}} &={-}\tfrac{2\varrho''}{\varrho}\bs{\dd} \rho \bs{\dd} \rho{+}\big[1{-}(\varrho\varrho')'\big](\bs{\dd} \vartheta \bs{\dd} \vartheta{+}\sin^2\vartheta\,\bs{\dd} \varphi \bs{\dd} \varphi)\;, 
    \\
    R_{\textrm{sym}} &=\tfrac{2}{\varrho^2} \left(1-2 \varrho \varrho''-\varrho'^2\right)\;.
\end{aligned}
\end{equation}

Let us study the weighted Li--Yau estimate \eqref{eq:LiYaugen} in spherically symmetric weighted spaces. As mentioned above, we first need to satisfy its unweighted version (with $w=1$). Recalling that it is met for spaces with non-negative Ricci curvature (and taking into account \eqref{eq:condpocatek}), we may find the constraints on the first and second derivatives,
\begin{equation}\label{eq:varrho}
    0\leq\varrho'\leq1\;, \quad \varrho''\leq0\;,\quad \varrho(0)=\varrho''(0)=0\;, \quad \varrho'(0)=1\;.
\end{equation}
Consequently, in any space $(\mathbb{R}^3,\bs{g}_{\textrm{sym}})$ with $\varrho$ given by \eqref{eq:varrho}, we can use the unweighted Li--Yau estimate. Following \eqref{eq:weightcond} (and considering \eqref{eq:condpocatek}), the heat kernel in the corresponding weighted spaces $(\mathbb{R}^3,\bs{g}_{\textrm{sym}},\mathfrak{s})$ can be estimated by the weighted Li--Yau estimate, if we choose the weight function $s$ such that
\begin{equation}\label{eq:weightconditiocna}
\begin{gathered}
    \int\limits_c^{\rho} \frac{d\tilde{\rho}}{\tilde{\rho}}s(\tilde{\rho}) V(\tilde{\rho}) \simeq s(\rho) V(\rho)\;, 
    \quad  \forall\rho>2c>0\;, 
    \\
    s(0)=1\;, \quad s'(0)=0\;,
\end{gathered}
\end{equation}
where ${V(\rho)=4\pi\int_0^{\rho} \!d\tilde{\rho}\,\varrho^2(\tilde{\rho})}$. The constraints in \eqref{eq:weightconditiocna} still offer many viable weight functions $s$. As a particular example, we can mention ${s(\rho)=C\rho^{\gamma}/V(\rho)}$ for ${\rho>2c}$ with ${C,\gamma>0}$ with an appropriate extension to ${\rho=0}$.

Great simplification can be achieved if the spherical symmetry is further incorporated into the problem. Specifically, if the source $J$ of our non-local equation is proportional to $\delta_{\mathrm{o}}$ or if we only need the values of field at the origin $\phi(\mathrm{o})$, then it is sufficient to know just the heat kernel with one point centered at the origin $\mathrm{o}$. In these situations, we can use the heat kernel estimate
\begin{equation}\label{eq:sphsymest}
    K_{\tau}^{\textrm{sym}}(\mathrm{x},\mathrm{o}) \asymp \frac{\tilde{c}}{V_{\mathfrak{s}}(\sqrt{\tau})} \exp \left(-c \frac{|\mathrm{x}|^2}{4\tau}\right)\;,
\end{equation}
which correctly approximates the heat kernel whenever \cite{Grigor:weight}
\begin{equation}\label{eq:VrS}
    V_{\mathfrak{s}}(\rho)\simeq\rho S_{\mathfrak{s}}(\rho)
\end{equation}
holds for large $\rho$. Remark that \eqref{eq:VrS} is always satisfied for a bounded range of $\rho$. Furthermore, if also ${V(\rho)\simeq\rho S(\rho)}$ for sufficiently large $\rho$ then necessarily ${V_{\mathfrak{s}}(\rho)\simeq s(\rho)V(\rho)}$, ${\forall\rho>0}$.  Let us stress that \eqref{eq:VrS} is the only the condition one has to satisfy. Therefore, it may hold with arbitrary Ricci curvature and even in the situations when the Li--Yau estimate is violated.

\subsection{Mean curvature}

The spherically symmetric weighted spaces described above can actually provide us with the heat kernel estimates even for spaces that are not spherically symmetric. Consider an arbitrary non-compact weighted manifold $(\mathcal{M},\bs{g},\mathfrak{w})$ with an origin ${\mathrm{o}\in\mathcal{M}}$ whose cut locus is an empty set. In any such a space, we can always introduce the spherical coordinates centered at $\mathrm{o}$ that cover the entire $\mathcal{M}\setminus\{\mathrm{o}\}$. The weighted Laplace operator in such coordinates reads
\begin{equation}\label{eq:trianglew}
    \triangle_{\mathfrak{w}}=\partial_{\rho}^2+M(\rho,\vartheta,\varphi)\partial_{\rho}+{\centernot\triangle} \;,
\end{equation}
where $M$ is the \textit{(weighted) mean curvature} of the geodesic 2-sphere $\partial\mathcal{B}(\mathrm{o},\rho)$ in the radial direction and ${\centernot\triangle}$ denotes the Laplace operator on $\partial\mathcal{B}(\mathrm{o},\rho)$. In the spherically symmetric weighted spaces ${(\mathbb{R}^3,\bs{g}_{\mathrm{sym}},\mathfrak{s})}$, we can write explicitly
\begin{equation}\label{eq:Msym}
\begin{aligned}
    M_{\textrm{sym}} &=\frac{S_{\mathfrak{s}}'}{S_{\mathfrak{s}}}=2\frac{\varrho'}{\varrho}+ \frac{s'}{s}\;, 
    \\
    {\centernot\triangle}_{\textrm{sym}} &=\frac{1}{\varrho^2}\left(\partial_{\vartheta}^2+\cot \vartheta\,\partial_{\vartheta} +\csc^2 \vartheta\, \partial_{\varphi}^{2} \right)\;.
\end{aligned}
\end{equation}

Let us assume that we find two spherically symmetric weighted spaces $(\mathbb{R}^3,\bs{g}_{\mathrm{sym}}^{\pm},\mathfrak{s}^{\pm})$ such that their mean curvatures $M^{\pm}$ delimit the mean curvature $M$ of  $(\mathcal{M},\bs{g},\mathfrak{w})$,
\begin{equation}\label{eq:MMM}
    M^{-}_{\textrm{sym}}(\rho)\geq M(\rho,\vartheta,\varphi)\geq M^{+}_{\textrm{sym}}(\rho)\;,
\end{equation}
with the identification of points  ${\mathrm{x}\in\mathcal{M}}$, ${\mathrm{x}^{\pm}\in \mathbb{R}^3}$ such that ${|\mathrm{x}|=|\mathrm{x}^{\pm}|}$. Then the heat kernels with one point centered at $\mathrm{o}$ satisfy inequalities \cite{Grigor:weight}
\begin{equation}\label{eq:KKK}
    K_{\tau}^{\textrm{sym}-}(\mathrm{x}^-,\mathrm{o}^-)\leq K_{\tau}(\mathrm{x},\mathrm{o})\leq K_{\tau}^{\textrm{sym}+}(\mathrm{x}^+,\mathrm{o}^+)\;.
\end{equation}
Combining with \eqref{eq:sphsymest}, these inequalities give rise to the estimate of $K_{\tau}(\mathrm{x},\mathrm{o})$. In fact, the relation \eqref{eq:KKK} may even remedy shortcomings of the above heat kernel estimates, in particular, the unspecified constants $c$ and $\tilde{c}$. This issue can be overcome whenever $K_{\tau}^{\textrm{sym}\pm}$ are known exactly (e.g., $K_{\tau}^{\textrm{euc}}$ or $K_{\tau}^{\textrm{hyp}}$).

%%%%%%%%%%%%%%%%%%%%%%%%%%%%%%%%%%%%%%%%%%%%%%%%%%%%%%%%%%%%%%%%%%%%%%%%%%%%%%%%
%%%%%%%%%%%%%%%%%%%%%%%%%%%%%%%%%%%%%%%%%%%%%%%%%%%%%%%%%%%%%%%%%%%%%%%%%%%%%%%%
%% Static Green's functions

\section{Static Green's functions} \label{sec:staticgreensfunctions}

We are now well equipped to solve the non-local Poisson's field equation for static scalar fields in static spacetimes \eqref{eq:redeq1},
\begin{equation*}
    \boxed{
    -\triangle_{\mathfrak{w}}\phi=e^{\ell^2\triangle_{\mathfrak{w}}}J\;.}
\end{equation*}
We will employ the method of Green's functions, which can be calculated or at least estimated directly from the heat kernels and their estimates. 

\subsection{Static Green's functions from heat kernels}

Starting with the non-compact manifold, we define the \textit{static Green's function} ${G(\mathrm{x},\mathrm{y})}$ as a solution of the non-local equation above with the point source ${J=\delta_{\mathrm{y}}}$ satisfying the boundary condition ${G\to0}$ for ${D_{\mathrm{xy}}\to\infty}$. Employing \eqref{eq:Kjeenadelta}, we may write 
\begin{equation}
    -\triangle_{\mathfrak{w}}G(\cdot,\mathrm{y})=e^{\ell^2\triangle_{\mathfrak{w}}}\delta_{\mathrm{y}}=K_{\ell^2}(\cdot,\mathrm{y})\;.
\end{equation}
The solution with an arbitrary source ${J\in F\mathcal{M}}$ is then given by
\begin{equation}
    \phi(\mathrm{x})=\langle G(\mathrm{x},\cdot),J\rangle_{\mathfrak{w}}\;.
\end{equation}
The static Green's function $G$ can be calculated from its local version $G_{\textrm{loc}}$ through
\begin{equation}\label{eq:GGloc}
    G =e^{\ell^2\triangle_{\mathfrak{w}}}  G_{\textrm{loc}}\;,
    \quad
    -\triangle_{\mathfrak{w}}G_{\textrm{loc}}(\cdot,\mathrm{y})=\delta_{\mathrm{y}}\;.
\end{equation}

Of the utmost importance, however, is the following formula relating $G$ to an integral of $K_{\tau}$,\footnote{To the best of our knowledge, in the context of non-local fields, this formula was first used in \cite{Frolov:2015bia}.}
\begin{equation}\label{eq:GreenusingKT}
    G(\mathrm{x},\mathrm{y})=\int\limits_{\ell^2}^{\infty} \!d\tau\, K_{\tau}(\mathrm{x},\mathrm{y})\;.
\end{equation}
This integral expression can be even considered as an alternative definition of $G$, as it is often common for $G_{\textrm{loc}}$ \cite{Grigor:weight}, which is recovered by taking ${\ell\to0}$. The equation \eqref{eq:GGloc} is regained by means of the identity \eqref{eq:semigroup2}. The integral relation \eqref{eq:GreenusingKT} can be proven formally by a direct calculation
\begin{equation}
    \triangle_{\mathfrak{w}}G =\int\limits_{\ell^2}^{\infty} \!d\tau \,\triangle_{\mathfrak{w}} K_{\tau}
    =\int\limits_{\ell^2}^{\infty} \!d\tau \,\partial_{\tau}K_{\tau}
    =\left[K_{\tau}\right]_{\ell^2}^{\infty}=-K_{\ell^2}\;,
\end{equation}
where we used ${K_{\tau}\to 0}$ for ${\tau\to \infty}$. This condition is satisfied, for example, in the weighted manifolds admitting the Li--Yau estimate \eqref{eq:LiYaugen}. The convergence of the integral \eqref{eq:GreenusingKT} will be discussed in the next subsection. Here, we just mention that the local static Green's function  $G_{\textrm{loc}}(\mathrm{x},\mathrm{y})$ diverges at coinciding points ${\mathrm{x}=\mathrm{y}}$ due to asymptotic behavior for small $\tau$ \eqref{eq:HKtau0approx}.

The definition of the static Green's functions on compacts manifolds requires a modification. To understand why, let us apply the divergence theorem to the non-local equation \eqref{eq:redeq1}. We observe that the source $J$ integrated over the entire manifold $\mathcal{M}$ must vanish,
\begin{equation}\label{eq:divtheo}
    \int\limits_{\mathcal{M}} \mathfrak{w} J =-\int\limits_{\mathcal{M}} {\di}\big(\mathfrak{w}\,{\bs\dd}^{\sharp}e^{-\ell^2\triangle_{\mathfrak{w}}}\phi\big)=0\;,
\end{equation}
because ${\partial\mathcal{M}=\emptyset}$. This statement can be equivalently reformulated as ${J_0=\langle J,\psi_{0}\rangle_{\mathfrak{w}}=0}$ thanks to \eqref{eq:firsteigen}, where $J_k$ denotes the coefficients calculated with the help of \eqref{eq:expansionoff}. Since ${\langle\delta_{\mathrm{y}},1\rangle_{\mathfrak{w}}=1\neq 0}$, we introduce the static Green's function as the solution with the source ${J=\delta_{\mathrm{y}}-{1}/{V_{\mathfrak{w}}^{\mathcal{M}}}}$ (see, e.g., \cite{Aubin1998,Chapling2016} for the local case). Using \eqref{eq:Kjeenadelta}, and the spectral properties \eqref{eq:firsteigen}, \eqref{eq:infsumeig}, and \eqref{eq:heatkerneleigen}, we may express it equivalently as\footnote{The existence of smooth solutions on compact Riemannian manifolds was proven in \cite{Grka2012v}.}
\begin{equation}\label{eq:compGF}
\begin{aligned}
    -\triangle_{\mathfrak{w}}G(\cdot,\mathrm{y}) &=e^{\ell^2\triangle_{\mathfrak{w}}}\left(\delta_{\mathrm{y}}-\frac{1}{V_{\mathfrak{w}}^{\mathcal{M}}}\right)  =K_{\ell^2}(\cdot,\mathrm{y})-\frac{1}{V_{\mathfrak{w}}^{\mathcal{M}}} 
    \\
    &=e^{\ell^2\triangle_{\mathfrak{w}}}\left(\sum_{k=1}^{\infty}\psi_{k}(\cdot)\psi_{k}(\mathrm{y})\right) 
    \\
    &=\sum_{k=1}^{\infty}e^{-\ell^2\lambda_k}\psi_{k}(\cdot)\psi_{k}(\mathrm{y})\;,
\end{aligned}
\end{equation}
where the sums run from ${k=0}$ (as a consequence of ${J_0=0}$). The static Green's function contains a freedom in an arbitrary additive constant, which we will denote by ${C_0}$. This follows again from the fact that the only harmonic function (homogeneous solution) on a compact manifold is a constant function. The solution for an arbitrary source ${J\in F\mathcal{M}}$ satisfying \eqref{eq:divtheo}, ${J_0=0}$, is then still given by
\begin{equation}
    \phi(\mathrm{x})=\langle G(\mathrm{x},\cdot),J\rangle_{\mathfrak{w}}\;.
\end{equation}
The relation to its local counterpart now reads
\begin{equation}
    G =e^{\ell^2\triangle_{\mathfrak{w}}}  G_{\textrm{loc}}\;,
    \;\;
    -\triangle_{\mathfrak{w}}G_{\textrm{loc}}(\cdot,\mathrm{y})=\delta_{\mathrm{y}}-\frac{1}{V_{\mathfrak{w}}^{\mathcal{M}}}\;.
\end{equation}

The integral formula \eqref{eq:GreenusingKT} must be also modified by subtracting the inverse volume (corresponding to the limit ${K_{\tau}\to{1}/{V_{\mathfrak{w}}^{\mathcal{M}}}}$, for ${\tau\to\infty}$),
\begin{equation}\label{eq:gfcompact}
\begin{aligned}
    G(\mathrm{x},\mathrm{y}) &=\int\limits_{\ell^2}^{\infty} \!d\tau\, \left(K_{\tau}(\mathrm{x},\mathrm{y})-\frac{1}{V_{\mathfrak{w}}^{\mathcal{M}}} \right)+C_0
    \\
    &=\sum_{k=1}^{\infty}\frac{e^{-\ell^2\lambda_k}}{\lambda_k}\psi_{k}(\mathrm{x})\psi_{k}(\mathrm{y})+C_0\;,
\end{aligned}
\end{equation}
which can be alternatively derived by means of the spectral expansion of the field equation. It remains true also in the local case, ${\ell\to0}$. Divergence of $G_{\textrm{loc}}(\mathrm{x},\mathrm{y})$ for the coinciding points ${\mathrm{x}=\mathrm{y}}$ is still a consequence of \eqref{eq:HKtau0approx}. On the other hand, the non-local static Green's function $G(\mathrm{x},\mathrm{y})$, converges absolutely and uniformly ${\forall\mathrm{x},\mathrm{y}\in\mathcal{M}}$. This follows from the relation to the majorant $m_{k}(\ell^2)$ in the convergence of ${K_{\ell^2}(\mathrm{x},\mathrm{y})}$, because
\begin{equation}
    \sup_{\mathrm{x},\mathrm{y}\in\mathcal{M}}\Big|\tfrac{e^{-\ell^2\lambda_k}}{\lambda_k}\psi_{k}(\mathrm{x})\psi_{k}(\mathrm{y})\Big|\leq\frac{1}{\lambda_1}m_{k}(\ell^2)\;.
\end{equation}

\subsection{Estimates and regularity}

Returning to the non-compact case, we can study the regularity of \eqref{eq:GreenusingKT} using the Li--Yau estimate leading to
\begin{equation}\label{eq:Gestim}
    G(\mathrm{x},\mathrm{y}) \asymp \int\limits_{\ell^2}^{\infty} \!d\tau\,\frac{c}{V_{\mathfrak{w}}(\mathrm{x}, \sqrt{\tau})} \exp \left(-\tilde{c} \frac{D_{\mathrm{xy}}^2}{4\tau}\right)\;,
\end{equation}
which we refer to as the \textit{Gaussian estimate of $G$}. As we will see shortly, it is possible to derive an even simpler estimate for the static Green's function. If ${D_{\mathrm{xy}}\geq\ell}$, then we may write
\begin{equation}
    G(\mathrm{x},\mathrm{y}) \asymp\Bigg[\!\int\limits_{\ell^2}^{D_{\mathrm{xy}}^2}\!+\!\!\int\limits_{D_{\mathrm{xy}}^2}^{\infty}\Bigg]d\tau \frac{c e^{-\tilde{c} \frac{D_{\mathrm{xy}}^2}{4\tau}}}{V_{\mathfrak{w}}(\mathrm{x}, \sqrt{\tau})} \begin{cases}\leq\int\limits_{D_{\mathrm{xy}}^2}^{\infty}\!\!d\tau \frac{\frac13 c_3c_2+c}{V_{\mathfrak{w}}(\mathrm{x}, \sqrt{\tau})}\;,
    \\
    \geq\int\limits_{D_{\mathrm{xy}}^2}^{\infty}\!\!d\tau \frac{ce^{-\tilde{c}/4}}{V_{\mathfrak{w}}(\mathrm{x}, \sqrt{\tau})}\;,
    \end{cases}\!
\end{equation}
The upper bound of the first integral is a consequence of the volume doubling \eqref{eq:voldoub} and the reverse volume doubling~\eqref{eq:voldoubcons},
\begin{equation}
\begin{aligned}
    \int\limits_{\ell^2}^{\mathclap{D_{\mathrm{xy}}^2}}\!d\tau \tfrac{c e^{-\tilde{c} \frac{D_{\mathrm{xy}}^2}{4\tau}}}{V_{\mathfrak{w}}(\mathrm{x}, \sqrt{\tau})} &\leq \int\limits_{0}^{\mathclap{D_{\mathrm{xy}}^2}}\!d\tau \tfrac{c e^{-\tilde{c} \frac{D_{\mathrm{xy}}^2}{4\tau}}}{V_{\mathfrak{w}}(\mathrm{x}, \sqrt{\tau})} \leq \tfrac{c c_1\,D_{\mathrm{xy}}^{2\kappa}}{V_{\mathfrak{w}}(\mathrm{x}, D_{\mathrm{xy}})}\int\limits_{0}^{\mathclap{D_{\mathrm{xy}}^2}}\!d\tau \tfrac{e^{-\tilde{c} \frac{D_{\mathrm{xy}}^2}{4\tau}}}{\tau^{\kappa}}
    \\
    &=\tfrac{c_2D_{\mathrm{xy}}^2}{V_{\mathfrak{w}}(\mathrm{x},D_{\mathrm{xy}})}\leq \tfrac{c_3c_2D_{\mathrm{xy}}^2}{V_{\mathfrak{w}}(\mathrm{x},2D_{\mathrm{xy}})}
    \\
    &\leq\int\limits_{D_{\mathrm{xy}}^2}^{\mathclap{4D_{\mathrm{xy}}^2}}\!\!d\tau\tfrac{\tfrac13 c_3c_2}{V_{\mathfrak{w}}(\mathrm{x},\sqrt{\tau})}\leq\int\limits_{D_{\mathrm{xy}}^2}^{\infty}\!\!d\tau\tfrac{\tfrac13 c_3c_2}{V_{\mathfrak{w}}(\mathrm{x},\sqrt{\tau})}\;,
\end{aligned}
\end{equation}
where we denoted $c_2:=c c_1  \E_{2-\kappa}\left(\tilde{c}/4\right)>0$. (Here, the letter $\E$ stands for the exponential integral function.) On the other hand, if ${D_{\mathrm{xy}}<\ell}$, then we obtain
\begin{equation}
    G(\mathrm{x},\mathrm{y})\asymp \int\limits_{\ell^2}^{\infty} \!d\tau\frac{c e^{-\tilde{c} \frac{D_{\mathrm{xy}}^2}{4\tau}}}{V_{\mathfrak{w}}(\mathrm{x}, \sqrt{\tau})} 
    \begin{cases} \leq\int\limits_{\ell^2}^{\infty} \!d\tau\frac{c}{V_{\mathfrak{w}}(\mathrm{x}, \sqrt{\tau})}\;,
    \\
    \geq\int\limits_{\ell^2}^{\infty} \!d\tau\frac{c e^{-\tilde{c}/4}}{V_{\mathfrak{w}}(\mathrm{x}, \sqrt{\tau})} \;.
    \end{cases}
\end{equation}
Putting everything together, we get
\begin{equation}\label{eq:GFestimate}
    G(\mathrm{x},\mathrm{y})\simeq \int\limits_{\mathclap{\max^2\left(\ell,D_{\mathrm{xy}}\right)}}^{\infty} \!\frac{d\tau}{V_{\mathfrak{w}}(\mathrm{x}, \sqrt{\tau})}\;,
\end{equation}
which generalizes the known formula for the estimate of $G_{\textrm{loc}}$ from \cite{Grigor:weight}, where the bottom limit becomes $D_{\mathrm{xy}}^2$ in the limit~${\ell\to0}$. We call this expression the \textit{volume estimate of $G$}.

Based on \eqref{eq:GFestimate}, we can now discuss the regularity of the static Green's function. We observe that $G$ converges whenever the upper limit is finite,
\begin{equation}\label{eq:localGFfinite}
    \int\limits^{\infty} \!\frac{d\tau}{V_{\mathfrak{w}}(\mathrm{x}, \sqrt{\tau})}<\infty\;,
\end{equation}
and the bottom limit is strictly positive
\begin{equation}\label{eq:maxcond}
    \max\left(\ell,D_{\mathrm{xy}}\right)>0\;,
\end{equation}
because $V_{\mathfrak{w}}(\mathrm{x}, \sqrt{\tau})$ is a positive non-decreasing function of $\tau$ that vanishes for ${\tau=0}$. Note that only the condition \eqref{eq:maxcond} depends on the scale of non-locality, but \eqref{eq:localGFfinite} is independent of $\ell$. This reflects the fact that the non-locality introduced through the exponential operator affects only the short-distance behavior, while the long-distance asymptotic behavior far from the source remain the same.

The condition \eqref{eq:maxcond} is satisfied for $G_{\textrm{loc}}$ only if ${\mathrm{x}\neq \mathrm{y}}$. As mentioned above, $G_{\textrm{loc}}$ diverges at the coinciding points ${\mathrm{x}=\mathrm{y}}$. Of course, this is not the true for the non-local case where \eqref{eq:maxcond} is always satisfied since the scale of non-locality is positive, ${\ell>0}$. Indeed, the value ${\ell^2}$ plays a role of the bottom cutoff in \eqref{eq:GFestimate} (or \eqref{eq:Gestim}); it effectively regularizes the limit of coinciding points by $\int_{\ell^2}^{\infty}\!{d\tau}/{V_{\mathfrak{w}}(\mathrm{x}, \sqrt{\tau})}<\infty$ provided that \eqref{eq:localGFfinite} is met.

The condition \eqref{eq:localGFfinite} is related to the weighted volume growth. Using the limit comparison test for improper integrals, we can find that it is satisfied, for instance, if ${V_{\mathfrak{w}}(\mathrm{x}, \rho)\simeq \rho^{\eta}}$, ${\eta>2}$, for large $\rho$. This is still in accordance with the Li--Yau estimate if we take ${\eta=\alpha+\beta>2}$ above.

Let us consider a spherically symmetric weighted space ${(\mathbb{R}^3,\bs{g}_{\textrm{sym}},\mathfrak{s})}$ with the Dirac delta point source located at the origin, i.e., ${\delta_{\mathrm{o}}}$. The solution is then described by $G^{\textrm{sym}}(\mathrm{x},\mathrm{o})$. Following the same steps as before, but starting with \eqref{eq:sphsymest}, we can arrive at the Gaussian estimates of $G^{\textrm{sym}}$ centered at $\mathrm{o}$,
\begin{equation}\label{eq:Gsym1}
    G^{\textrm{sym}}(\mathrm{x},\mathrm{o})\asymp \int\limits_{\ell^2}^{\infty}\! d\tau\,
    \frac{\tilde{c}}{V_{\mathfrak{s}}(\sqrt{\tau})} \exp \left(-c \frac{|\mathrm{x}|^2}{4\tau}\right)\;,
\end{equation}
and the volume estimates of $G^{\textrm{sym}}$ centered at $\mathrm{o}$,
\begin{equation}\label{eq:Gsym2}
    G^{\textrm{sym}}(\mathrm{x},\mathrm{o})\simeq\int\limits_{\mathclap{\max^2\left(\ell,|\mathrm{x}|\right)}}^{\infty} \!\frac{d\tau}{V_{\mathfrak{s}}(\sqrt{\tau})}\;.
\end{equation}
Remark that the latter requires the volume doubling property ${V_{\mathfrak{s}}(2\rho)\leq c V_{\mathfrak{s}}(\rho)}$, ${\forall\rho>0}$, to hold.

The estimates \eqref{eq:Gsym1} and \eqref{eq:Gsym2} can be contrasted with the exact formula for the static Green's function center at $\mathrm{o}$ that is available in the local case,
\begin{equation}\label{eq:Gsymloc}
    G_{\textrm{loc}}^{\textrm{sym}}(\mathrm{x},\mathrm{o})=\int\limits_{|\mathrm{x}|}^{\infty}\frac{d\rho}{S_{\mathfrak{s}}(\rho)}\;.
\end{equation}
We may verify \eqref{eq:Gsymloc} as follows:
\begin{equation}
\begin{aligned}
    \langle -\triangle_{\mathfrak{s}}G_{\textrm{loc}}^{\textrm{sym}}(\cdot,\mathrm{o}),f\rangle_{\mathfrak{s}}&=\langle \partial_{\rho}G_{\textrm{loc}}^{\textrm{sym}}(\cdot,\mathrm{o}) ,\partial_{\rho}f\rangle_{\mathfrak{s}}=-\!\int\limits_{\mathcal{M}} \frac{\mathfrak{s}\partial_{\rho}f}{S_{\mathfrak{s}}}
    \\
    &=-\!\int\limits_{0}^{\infty}\!d\rho\; \partial_{\rho}f=f(0)=\langle \delta_{\mathrm{o}},f\rangle_{\mathfrak{s}}\;.
\end{aligned}
\end{equation}

As before, we still get the convergence condition ${\eta=\alpha+\beta>2}$ for the upper limit of \eqref{eq:Gsym2} (or \eqref{eq:Gsym1} with ${\mathrm{x}=\mathrm{o}}$) where ${\varrho\simeq\rho^{(\alpha-1)/2}}$ and ${s(\rho)\simeq \rho^{\beta}}$. Calculating the Riemann tensor of the full 4-dimensional Lorentzian manifold $(\mathbb{R}^4,\bar{\bs{g}}_{\textrm{sym}})$, we find that it vanishes asymptotically if \begin{equation}
    s=O(1)\;, \quad \varrho=\rho+O(1)\;, \quad \rho\to\infty\;.
\end{equation}
Consequently, an asymptotic flatness of the static symmetric spacetime is sufficient for the convergence of ${G^{\textrm{sym}}_{\textrm{loc}}(\mathrm{x},\mathrm{o})}$, ${\mathrm{x}\neq\mathrm{o}}$ and ${G^{\textrm{sym}}(\mathrm{x},\mathrm{o})}$, ${\forall\mathrm{x}\in\mathbb{R}^3}$.

Finally, let us also mention that the inequalities between the heat kernels \eqref{eq:KKK} can be promoted to the inequalities between the corresponding static Green's functions centered at $\mathrm{o}$,
\begin{equation}\label{eq:GGG}
    G^{\textrm{sym}-}(\mathrm{x}^-,\mathrm{o}^-)\leq G(\mathrm{x},\mathrm{o})\leq G^{\textrm{sym}+}(\mathrm{x}^+,\mathrm{o}^+)\;.
\end{equation}
We refer to them as the \textit{plus--minus estimate}. Recall that $G$ can be the Green's function on an arbitrary weighted manifold with an origin that is missing its cut locus, provided that \eqref{eq:MMM} are satisfied. Furthermore, the plus-minus estimate holds true even in the local case.

\subsection{Example: Ext. \& int. Schwarzschild spacetime}

As a simple example, consider a static spacetime composed of the \textit{exterior Schwarzschild} geometry glued to the \textit{interior Schwarzschild} metric \cite{Schwarzschild:1916ae}. In particular, we choose two constants ${0<\tfrac98 a<b}$, where ${r=b}$ is the matching surface, ${r=a}$ is the Schwarzschild radius, and ${r=0}$ is the origin $\mathrm{o}$.
Therefore, the metric for ${0<r<b}$ is
\begin{equation}
\begin{aligned}
    \bar{\bs{g}}_{\textrm{sch}}^{\textrm{int}} &= -\left(\tfrac{3}{2} \sqrt{1-\tfrac{a}{b}}-\tfrac{1}{2}\sqrt{1-\tfrac{r^{2} a}{b^{3}}}\right)^{2}\bs{\dd} T \bs{\dd} T+\frac{\bs{\dd} r \bs{\dd} r}{1-\frac{a}{b^3}r^2}
        \\
    &\feq+r^2(\bs{\dd} \vartheta \bs{\dd} \vartheta+\sin^2\vartheta\,\bs{\dd} \varphi \bs{\dd} \varphi)\;,
\end{aligned}
\end{equation}
while the metric for ${r> b}$ reads
\begin{equation}
\begin{aligned}
    \bar{\bs{g}}_{\textrm{sch}}^{\textrm{ext}} &= -\big(1-\tfrac{a}{r}\big)\,\bs{\dd} T \bs{\dd} T+\frac{\bs{\dd} r \bs{\dd} r}{1-\frac{a}{r}}
        \\
    &\feq+r^2(\bs{\dd} \vartheta \bs{\dd} \vartheta+\sin^2\vartheta\,\bs{\dd} \varphi \bs{\dd} \varphi)\;.
\end{aligned}
\end{equation}
Both geometries can be rewritten in the form
\begin{equation}
\begin{aligned}
    \bar{\bs{g}}_{\textrm{sch}} &= -s^2(\rho)\bs{\dd} t \bs{\dd} t+\bs{\dd} \rho \bs{\dd} \rho
    \\
    &\feq+\varrho^2(\rho)(\bs{\dd} \vartheta \bs{\dd} \vartheta+\sin^2\vartheta\,\bs{\dd} \varphi \bs{\dd} \varphi)\;,
\end{aligned}
\end{equation}
if we use the transformation of the radial coordinate
\begin{equation}
\begin{aligned}
    \rho &:=\xi(r^2) :=\begin{cases}
    \frac{b^{3/2}}{\sqrt{a}} \sin ^{-1}\left(\frac{ \sqrt{a}}{b^{3/2}}r\right)\;, & r< b\;,
    \\
    a \tanh ^{-1}\left(\sqrt{1-\tfrac{a}{r}}\right)
    \\
    +r \sqrt{1-\tfrac{a}{r}}+\xi_0\;, & r>b\;,
    \end{cases}
    \\
    \xi_0 &:=\tfrac{b^{3/2}}{\sqrt{a}} \sin ^{-1}\left(\tfrac{\sqrt{a}}{\sqrt{b}}\right)-b \sqrt{1-\tfrac{a}{b}}
    \\
    &\feq-a \tanh ^{-1}\left(\sqrt{1-\tfrac{a}{b}}\right)\;,
\end{aligned}
\end{equation}
and rescale the temporal coordinate as 
\begin{equation}
    t:=T/s_{\infty}\;, \quad s_{\infty}:=1/\left(\tfrac{3}{2} \sqrt{1-\tfrac{a}{b}}-\tfrac{1}{2}\right)\;.
\end{equation}
The constant $\xi_0$ was fixed from the continuity of the radial coordinate and the constant $s_{\infty}$ was chosen so that ${s(0)=1}$. As a result, we can view this spacetime as the spherically symmetric weighted space ${(\mathbb{R}^3,\bs{g}_{\textrm{sch}},\mathfrak{s})}$. The functions $\varrho$ and $s$ are given implicitly in terms of $\xi$,\footnote{The fact that $\varrho$ and $s$ are not twice differentiable is not problematic because $\triangle_{\mathfrak{s}}$ depends only on zeroth and first derivatives, see \eqref{eq:trianglew} and \eqref{eq:Msym}.}
\begin{equation}
\begin{aligned}
    \varrho^2(\rho) &=\xi^{-1}(\rho)\;,
\\
    s\big(\xi(r^2)\big) &=s_{\infty}\times\begin{cases}\tfrac{3}{2} \sqrt{1-\tfrac{a}{b}}-\tfrac{1}{2}\sqrt{1-\tfrac{a  }{b^{3}}r^2}\;, & r< b\;,
    \\
    \sqrt{1-\tfrac{a}{r}}\;, & r>b\;.\\
    \end{cases}
\end{aligned}
\end{equation}
Unfortunately, $\xi^{-1}$ does not have a closed-form expression for ${r>b}$. However, as we will see we can make a significant progress even without it. For instance, using the formula for the integral of an inverse function \cite{Laisant:1905} and employing ${\varrho(0)=0}$, we can calculate $V(\rho)$,
\begin{equation}
    \frac{V(\rho)}{4\pi}=\int\limits_0^{\rho}\!d \tilde{\rho} \; \xi^{-1}(\tilde{\rho}) =\rho \xi^{-1}(\rho)-\Xi(\xi^{-1}(\rho))+\Xi(0)\;,
\end{equation}
where we denoted the primitive function of $\xi$ by $\Xi$,
\begin{equation}
    \begin{aligned}
        \Xi(r^2) &=\begin{cases}
        \frac{r \sqrt{a \left(b^3-a r^2\right)}-\left(b^3-2 a r^2\right) \sin ^{-1}\left(r \sqrt{\frac{a}{b^3}}\right)}{2 \left(\frac{a}{b}\right)^{3/2}} \;, & r< b\;,
        \\
        \left(a r^2-\frac{5 a^3}{8}\right) \tanh ^{-1}\left(\sqrt{1-\frac{a}{r}}\right)
        \\
        +\left(-\tfrac{5}{8} a^2-\tfrac{5}{12} a r+\tfrac{2}{3} r^2\right) \sqrt{r (r-a)}
        \\
        +\xi_0 r^2+\Xi_0\;, &r>b\;,
        \end{cases}
        \\
        \Xi_0 &:=\tfrac{5}{8} a^3 \tanh ^{-1}\left(\sqrt{1-\tfrac{a}{b}}\right) 
        \\
        &\feq+\sqrt{b}\sqrt{b-a} \left(\tfrac58 a^2+\tfrac{ b^3}{2a}+\tfrac{5}{12} a b+\tfrac13 b^2\right)
        \\
        &\feq- \tfrac{b^{9/2}}{2a^{3/2}} \sin ^{-1}\left(\sqrt{\tfrac{a}{b}}\right)\;.
    \end{aligned}
\end{equation}
The integration constant was chosen so that ${\Xi(0)=0}$. 

Let us study the static Green's functions centered at the origin, ${r=\rho=0}$. The exact expression for the local one can be obtained from \eqref{eq:Gsymloc}, 
\begin{equation}\label{eq:locgfschw}
    G_{\mathrm{loc}}^{\textrm{sch}}(\mathrm{x},\mathrm{o})=\frac{1}{4\pi }\int\limits_{|\mathrm{x}|}^{\infty} \frac{d \rho}{s(\rho)\xi^{-1}(\rho)}=\frac{1}{2\pi }\int \limits_{\mathclap{\sqrt{\xi^{-1}(|\mathrm{x}|)}}}^{\infty}\!dr\, \frac{\xi'(r^2)}{s(\xi(r^2))\, r}\;,
\end{equation}
which clearly diverges at ${\mathrm{x}=\mathrm{o}}$, since the integrand behaves as ${{1}/{4 \pi  r^2}+O(1)}$ for ${r\to0}$.

To use the formulas for the estimates of non-local static Green's functions \eqref{eq:Gsym1} and \eqref{eq:Gsym2}, we have to check their assumptions. The space ${(\mathbb{R}^3,\bs{g}_{\textrm{sch}},\mathfrak{s})}$ approaches the Euclidean 3-space for ${\rho\to0}$ and the weighted Euclidean 3-space for ${\rho\to\infty}$ with constant weight function ${s_{\infty}}$. It follows from ${\rho S(\rho)/V(\rho)\to 3}$ and ${s(\rho)\to s_{\infty}}$ for ${\rho\to\infty}$ that \eqref{eq:VrS} is actually satisfied for both weighted and unweighted versions. Thus, we can write the Gaussian estimate as
\begin{equation}\label{eq:schgfGE}
\begin{aligned}
    G^{\textrm{sch}}(\mathrm{x},\mathrm{o}) &\asymp \frac{\tilde{c} }{3(4\pi)^{\frac32}}\int\limits_{\ell^2}^{\infty}\! d\tau\,
    \tfrac{\exp \left(- {c|\mathrm{x}|^2}/{4\tau}\right)}{s\left(\xi(r^2)\right)\left[\sqrt\tau \xi^{-1}(\sqrt\tau)-\Xi(\xi^{-1}(\sqrt\tau))\right]}
    \\
    &=\frac{\tilde{c}}{6\pi^{\frac32}}\int\limits_{\mathclap{\sqrt{\xi^{-1}(\ell)}}}^{\infty}\! dr\,
    \frac{r \xi(r^2)\xi'(r^2)\,\exp \left(- \frac{c|\mathrm{x}|^2}{4\xi^2(r^2)}\right)}{s\big(\xi(r^2)\big)\left[r^2\xi(r^2) -\Xi(r^2)\right]}\;,
\end{aligned}
\end{equation}
where we used the relation ${V_{\mathrm{s}}(\rho)\simeq s(\rho)V(\rho)}$. In the derivation, we performed a change of integration variable, ${\xi^{-1}(\sqrt\tau)=r^2}$, to bring the integral to a more tractable form, involving the inverse function in the bottom limit only. Furthermore, we used ${\xi^{-1}(\infty)=\infty}$ and the formula for the derivative of an inverse function, ${\tau'(r)=4r\xi(r^2)\xi'(r^2)}$. Similarly, we can also obtain the volume estimate,
\begin{equation}\label{eq:schgfVE}
\begin{aligned}
    G^{\textrm{sch}}(\mathrm{x},\mathrm{o}) &\simeq \frac{1}{24\pi}\int\limits_{\mathclap{\max^2\left(\ell,|\mathrm{x}|\right)}}^{\infty}
    \tfrac{d\tau}{s\left(\xi(r^2)\right)\left[\sqrt\tau \xi^{-1}(\sqrt\tau)-\Xi(\xi^{-1}(\sqrt\tau))\right]}
    \\
    &=\frac{1}{6\pi}\int\limits_{\mathclap{\sqrt{\xi^{-1}\left(\max\left(\ell,|\mathrm{x}|\right)\right)}}}^{\infty}\! dr\,
    \frac{r \xi(r^2)\xi'(r^2)}{s\big(\xi(r^2)\big)\left[r^2\xi(r^2) -\Xi(r^2)\right]}\;.
\end{aligned}
\end{equation}
Note that the volume doubling property is satisfied since ${V_{\mathfrak{s}}(2\rho)/V_{\mathfrak{s}}(\rho)}$ is a positive continuous function that approaches a constant number $8$ for ${\rho\to0}$ and ${\rho\to\infty}$. The overall constant in both estimates was rescaled to match the asymptotic of \eqref{eq:locgfschw} for large $\rho$ if all constants are set to one, ${c_{\textrm{b}}=c_{\textrm{u}}=\tilde{c}_{\textrm{b}}= \tilde{c}_{\textrm{u}}=c_{\textrm{m}}=1}$. This is motivated by the fact that the non-locality described by the exponential operator usually affects the static field only close to the source.

\begin{figure}
    \centering
    \includegraphics[width=\columnwidth]{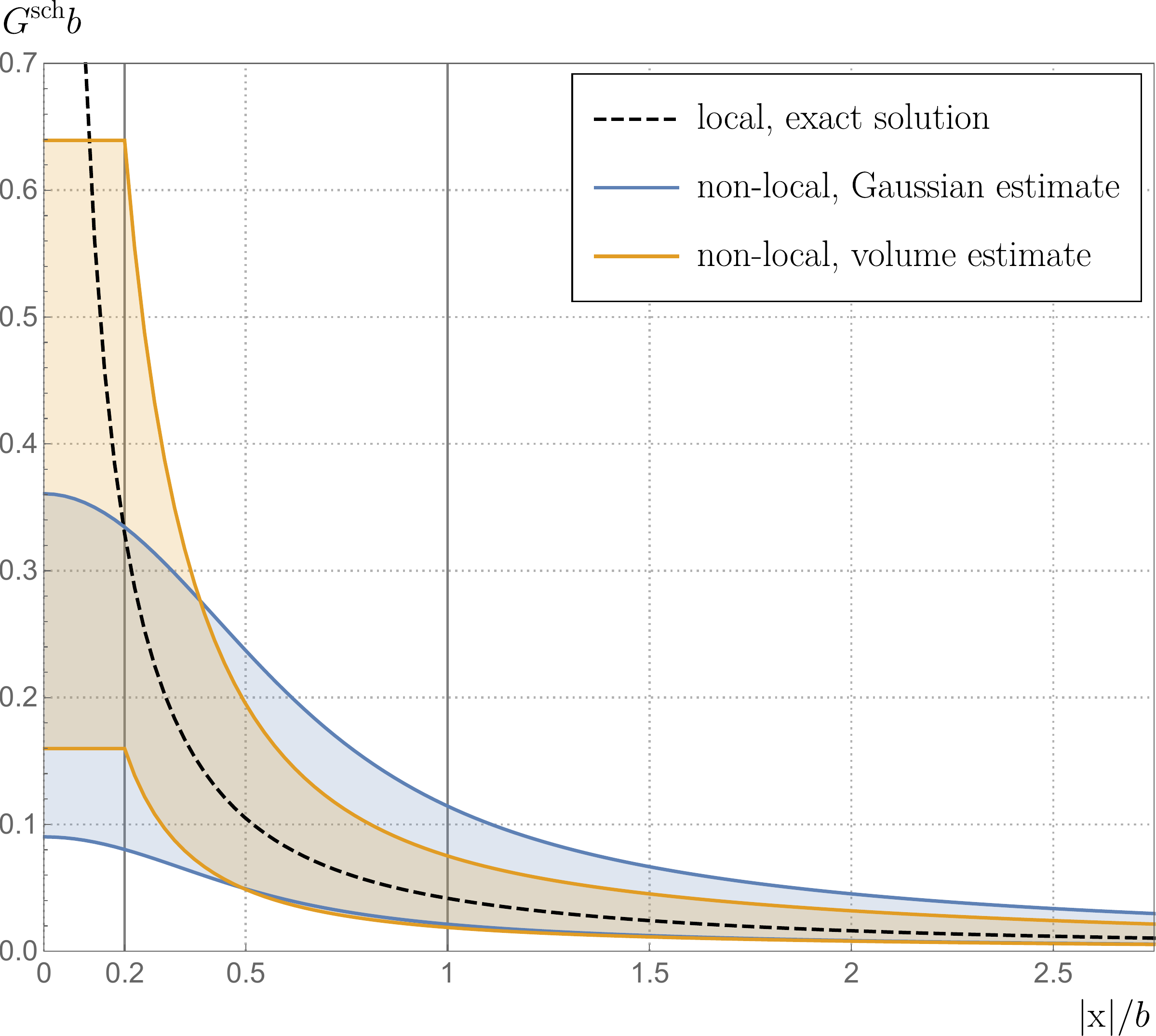}
    \caption{Static Green's functions centered at the origin calculated in a spacetime composed of exterior and interior Schwarzschild metrics. Dashed line is the exact local static Green's function given by \eqref{eq:locgfschw}. Blue region describes the Gaussian estimate of the non-local static Green's function \eqref{eq:schgfGE} with ${\tilde{c}_{\textrm{u}}=1/\tilde{c}_{\textrm{b}}=2}$ and ${c_{\textrm{u}}=1/c_{\textrm{b}}=4/5}$. The yellow region corresponds to the volume estimate of the non-local static Green's function \eqref{eq:schgfVE} with ${c_{\textrm{m}}=2}$. We set ${\ell/b=1/5}$ and ${a/b=7/9}$.}
    \label{fig:sch}
\end{figure}

Integral \eqref{eq:locgfschw} has a closed-form, which is rather lengthy and not very illuminating. Integrals \eqref{eq:schgfGE} and \eqref{eq:schgfVE} do not have a closed form, but they converge everywhere due to constantly weighted Euclidean asymptotic behavior of the integrand. They can be easily evaluated numerically, see Fig.~\ref{fig:sch}. Although the constants in the estimates of the non-local static Green's function are arbitrary, one can clearly see that the exact solution must be finite, which is not true for the local static Green's function at ${\mathrm{x}=\mathrm{o}}$. Let us point out that the estimate \eqref{eq:schgfVE} with ${\ell\to0}$ approximates the exact local solution \eqref{eq:locgfschw} rather well because ${{\Xi(r^2)/r^2\xi(r^2)}\approx 2/3}$. Finally, by taking the flat-space limit ${a\to0}$ of \eqref{eq:schgfGE} and \eqref{eq:schgfVE}, we get 
\begin{equation}
\begin{aligned}
    G^{\textrm{euc}}(\mathrm{x},\mathrm{o}) &\asymp\frac{\tilde{c} \erf\left(\frac{\sqrt{c} |x|}{2 \ell }\right)}{4 \pi  \sqrt{c} |x|}\;, 
    \\
    G^{\textrm{euc}}(\mathrm{x},\mathrm{o}) &\simeq\frac{1}{4 \pi  \max (|x|,\ell)}\;,
\end{aligned}
\end{equation}
which are estimates of the non-local static Green's function in the Euclidean 3-space \eqref{eq:gfeuc}. (Here, $\erf$ is the error function.) 

\subsection{Example: Ultrastatic universes}

Consider spacetimes of the homogeneous isotropic \textit{ultrastatic universes} of zero, negative, and positive spatial curvatures,
\begin{equation}
    \bar{\bs{g}}_{\textrm{uni}} := - \bs{\dd} t \bs{\dd} t+\begin{cases}
    \bs{g}_{\textrm{euc}}\;, &\mathcal{M}=\mathbb{R}^3\;,
    \\
    \bs{g}_{\textrm{hyp}}\;, &\mathcal{M}=\mathbb{R}^3\;,
    \\
    \bs{g}_{\textrm{sph}}\;, &\mathcal{M}=\mathbb{S}^3\;,
    \end{cases}
\end{equation}
where the 3-dimensional metrics are given by \eqref{eq:geuc}, \eqref{eq:ghyp}, and \eqref{eq:gsph}. They describe the \textit{Minkowski spacetime}, the \textit{hyperbolic universe}, and the \textit{Einstein universe}, respectively. 

Applying the formula \eqref{eq:GreenusingKT} to the heat kernel \eqref{eq:eucHK}, we can really easily calculate the static Green's function in the Minkowski spacetime (together with its local limit)
\begin{equation}\label{eq:gfeuc}
    G^{\textrm{euc}}(\mathrm{x},\mathrm{y}) =\frac{ \erf\left(\frac{D_{\mathrm{xy}}}{2 \ell }\right)}{4 \pi  D_{\mathrm{xy}}}\;, \quad    G^{\textrm{euc}}_{\textrm{loc}}(\mathrm{x},\mathrm{y}) =\frac{1}{4 \pi  D_{\mathrm{xy}}}\;,
\end{equation}
which was found earlier in \cite{Biswas:2011ar,Modesto:2010uh} using the Fourier transform and in \cite{Frolov:2015bia} by means of this heat kernel approach. 

Similarly, we can employ the formula \eqref{eq:GreenusingKT} together with \eqref{eq:Khyp} to find the static Green's function in the hyperbolic universe. Integrating the heat kernel we arrive at
\begin{equation}\label{eq:gfhyp}
\begin{aligned}
    G^{\textrm{hyp}}(\mathrm{x},\mathrm{y}) &=\tfrac{{D_{\mathrm{xy}}}/{A}}{\sinh \left({D_{\mathrm{xy}}}/{A}\right)}H_{D_{\mathrm{xy}},\ell}\left(1/A\right)\;, 
    \\
    G^{\textrm{hyp}}_{\textrm{loc}}(\mathrm{x},\mathrm{y}) &=\tfrac{{D_{\mathrm{xy}}}/{A}}{\sinh \left({D_{\mathrm{xy}}}/{A}\right)}H_{D_{\mathrm{xy}},0}\left(1/A\right)\;,
\end{aligned}
\end{equation}
where we denoted
\begin{equation}\label{eq:Hint}
\begin{aligned}
    H_{\alpha,\beta}(\gamma) &:=\int\limits_{\beta^2}^{\infty}\!d\tau \,\frac{e^{-\tau \gamma^2-\frac{\alpha^2}{4\tau}}}{(4\pi\tau)^{\frac32}}=\frac{1}{4 \pi  \alpha}\Big[\tfrac12 e^{-\gamma\alpha } \erf\big(\tfrac{\alpha}{2 \beta }{-}\gamma \beta \big)
    \\
    &\feq+\tfrac12 e^{\gamma \alpha} \erf\big(\tfrac{\alpha}{2 \beta }{+}\gamma\beta \big)-\sinh \left(\gamma \alpha\right)\Big]\;,
    \\
    H_{\alpha,0}(\gamma)&:=\frac{e^{-\gamma\alpha}}{4\pi \alpha}\;,
\end{aligned}
\end{equation}
for later convenience. The function \eqref{eq:gfhyp} goes to zero as $\sim e^{-{D_{\mathrm{xy}}}/{A}}$ for ${D_{\mathrm{xy}}\to\infty}$ in contrast to ${\sim 1/D_{\mathrm{xy}}}$ in the Euclidean space. 

Since the spatial part of the Einstein universe is compact, we have to use \eqref{eq:gfcompact} together with the addition theorem \eqref{eq:addtheo} to calculate the static Green's function. We obtain
\begin{equation}\label{eq:gfsph}
\begin{aligned}
    G^{\textrm{sph}}(\mathrm{x},\mathrm{y}) &= \sum_{k=1}^{\infty}\frac{e^{-k(k+2) \frac{\ell^2}{B^2}}}{2 \pi^{2}B}\frac{(k+1)^2}{k(k+2)}
    \\
    &\feq\times\left[\tfrac{\sin \left((k+1) \frac{D_{\mathrm{xy}}}{B}\right)}{(k+1)\sin\left( \frac{D_{\mathrm{xy}}}{B}\right)}-(-1)^k\right]\;,
    \\
    G^{\textrm{sph}}_{\textrm{loc}}(\mathrm{x},\mathrm{y}) &=\frac{\left(\pi -\frac{D_{\mathrm{xy}}}{B}\right) \cot \left(\frac{D_{\mathrm{xy}}}{B}\right)+1}{4 \pi ^2 B}\;,
\end{aligned}
\end{equation}
where the infinite sum converges absolutely and uniformly as we expected. In order to fix the conventional constant $C_0$ in \eqref{eq:gfcompact}, we employed a convention that the static Green's functions vanish for the longest geodesics, ${D_{\mathrm{xy}}=\pi B}$. The local expression matches known formula \cite{Szmytkowski2007,Chapling2016} found by other methods. 

\begin{figure}
    \centering
    \includegraphics[width=\columnwidth]{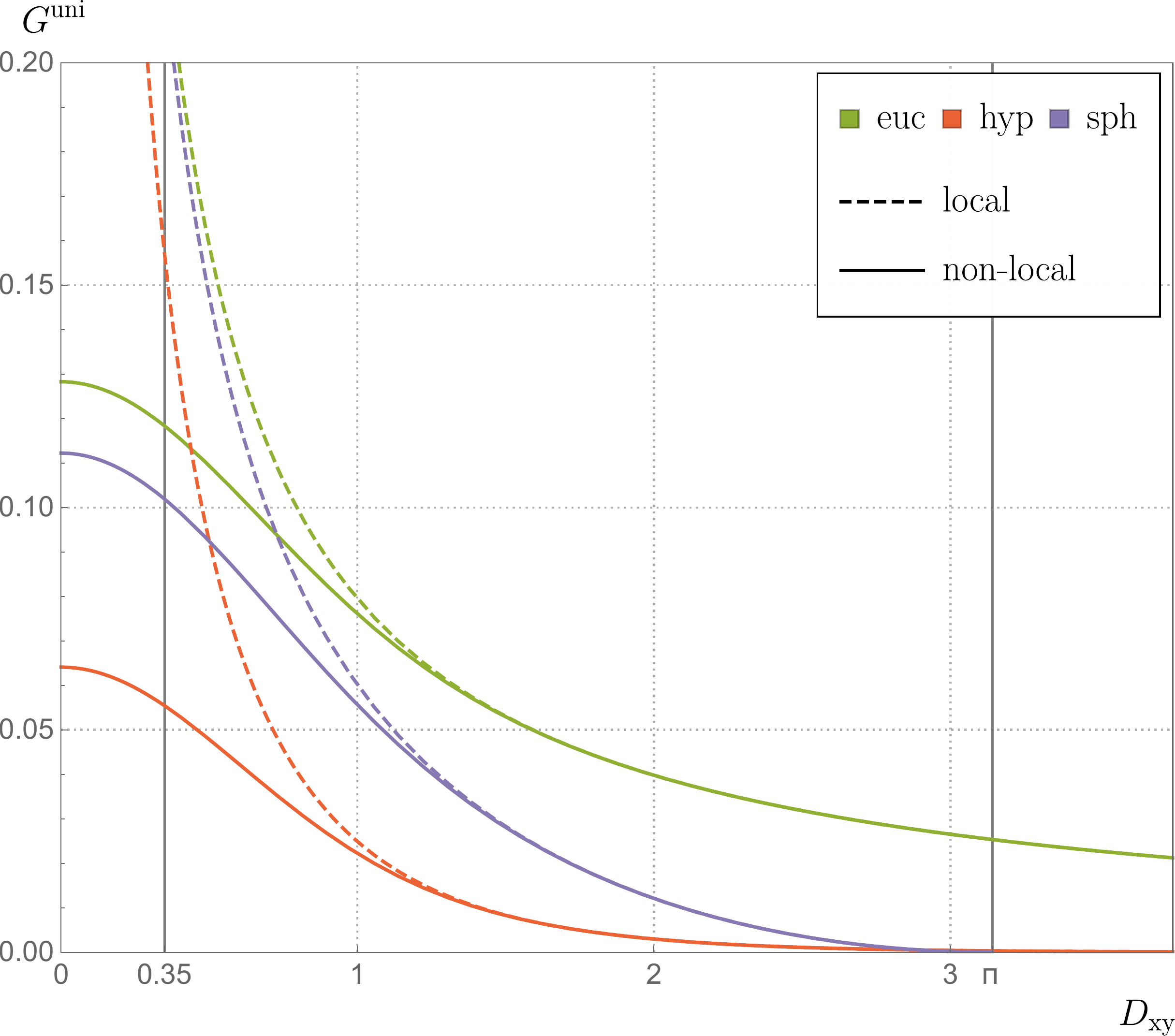}
    \caption{Static Green's functions calculated in ultrastatic universes: the Minkowski spacetime (euc), the hyperbolic universe (hyp), and the Einstein universe (sph). Dashed and solid lines describe the local and non-local static Green's function, respectively, see \eqref{eq:gfeuc}, \eqref{eq:gfhyp}, and \eqref{eq:gfsph}. We set ${\ell=0.35\,\textrm{m}}$ and ${A=B=1\,\textrm{m}}$ for better qualitative comparison. The end point ${D_{\mathrm{xy}}=\pi\,B}$ of the purple lines corresponds to longest geodesics between opposite poles on the sphere.}
    \label{fig:uni}
\end{figure}

In all three cases, the local Green's functions blow up at coinciding points ${\mathrm{x}=\mathrm{y}}$ while their non-local counterparts are finite everywhere. Moreover, \eqref{eq:gfhyp} and \eqref{eq:gfsph} reduce back to \eqref{eq:gfeuc} for large values of $A$ and $B$ (the latter we checked numerically in the non-local case). The graphs of the functions mentioned here are shown in Fig.~\ref{fig:uni}.

\subsection{Example: Anti-de Sitter spacetime}

Returning to non-compact manifolds, we would like to demonstrate that owing to \eqref{eq:GGG}, the result \eqref{eq:gfhyp} (and its special case \eqref{eq:gfeuc}) may actually serve as rather precise estimate for the static Green's function centered at the origin,
\begin{equation}\label{eq:estcomp}
    G^{\textrm{hyp}-}(\mathrm{x}^-,\mathrm{o}^-)\leq G(\mathrm{x},\mathrm{o})\leq G^{\textrm{hyp}+}(\mathrm{x}^+,\mathrm{o}^+)\;.
\end{equation}
Here, the two hyperbolic 3-spaces ${(\mathbb{R}^2,\bs{g}_{\textrm{hyp}}^{\pm})}$ differ by the constants ${A_{+}>A_{-}>0}$ (including the special case ${A_{+}=\infty}$ corresponding to the Euclidean 3-space $G^{\textrm{euc}+}$). This estimate is satisfied for an arbitrary weighted manifold ${(\mathcal{M},\bs{g},\mathfrak{w})}$ whose origin does not have a cut locus and which mean curvature $M$ satisfies \eqref{eq:MMM}. Compared to unspecified constants in previous estimates, the constants $A_{\pm}$ can be computed directly. 

Let us restrict ourselves to the spherically symmetric weighted space $(\mathbb{R}^3,\bs{g}_{\mathrm{sym}},\mathfrak{s})$. Then the condition \eqref{eq:MMM} takes the form
\begin{equation}
    \tfrac{2}{A_{-}} \coth \big(\tfrac{\rho }{A_{-}}\big)\geq \frac{S_{\mathfrak{s}}'(\rho)}{S_{\mathfrak{s}}(\rho)}\geq \tfrac{2}{A_{+}} \coth \big(\tfrac{\rho }{A_{+}}\big)\;.
\end{equation}
Interestingly, this differential inequality can be solved exactly using Gr\"onwall's lemma \cite{Gronwall}, which implies that
\begin{equation}
    S_{\mathfrak{s}}(\rho_0) \frac{\sinh ^2\big(\frac{\rho }{A_{-}}\big)}{\sinh ^2\big(\tfrac{\rho_0 }{A_{-}}\big)}\geq S_{\mathfrak{s}}(\rho)\geq S_{\mathfrak{s}}(\rho_0) \frac{\sinh ^2\big(\frac{\rho }{A_{+}}\big)}{\sinh ^2\big(\tfrac{\rho_0 }{A_{+}}\big)}\;,
\end{equation}
with ${\rho>\rho_0>0}$. Taking the limit ${\rho_{0}\to0}$, we obtain an inequality that holds for all ${\rho>0}$,
\begin{equation}
    4\pi A_{-}^2\sinh ^2\big(\tfrac{\rho }{A_{-}}\big)\geq S_{\mathfrak{s}}(\rho)\geq 4\pi A_{+}^2\sinh ^2\big(\tfrac{\rho }{A_{+}}\big)\;,
\end{equation}
where we used the fact that ${S_{\mathfrak{s}}(\rho_0)/4\pi\rho_0^2\to 1}$ for ${\rho_0\to0}$ as it follows from \eqref{eq:condpocatek}. Whenever ${S_{\mathfrak{s}}(\rho)=4\pi s(\rho)\varrho^2(\rho)}$ satisfies these inequalities, we can use the estimate \eqref{eq:estcomp}.

The previous considerations can be applied, for example, to the \textit{anti-de Sitter spacetime}. If we write it in the global coordinates,
\begin{equation}
\begin{aligned}
    \bar{\bs{g}}_{\textrm{ads}} &= -\cosh^2\big(\tfrac{\rho}{A_0}\big)\bs{\dd} t \bs{\dd} t+\bs{\dd} \rho \bs{\dd} \rho
    \\
    &\feq+A_0^2\sinh^2\big(\tfrac{\rho}{A_0}\big)(\bs{\dd} \vartheta \bs{\dd} \vartheta+\sin^2\vartheta\,\bs{\dd} \varphi \bs{\dd} \varphi)\;,
\end{aligned}
\end{equation}
we can immediately see that it can be represented as the weighted hyperbolic 3-space $(\mathbb{R}^3,\bs{g}_{\textrm{hyp}},\mathfrak{s})$ with the weight function ${s(\rho)=\cosh({\rho}/{A_0})}$. The weighted volume and surface of geodesic balls are given by
\begin{equation}
\begin{aligned}
    S_{\mathfrak{s}}(\rho) &= 4\pi A_0^2\sinh^2\big(\tfrac{\rho}{A_0}\big)\cosh\big(\tfrac{\rho}{A_0}\big)\;,
    \\ 
    V_{\mathfrak{s}}(\rho) &=\frac{4\pi}{3} A_0^3 \sinh ^3\big(\tfrac{\rho}{A_0}\big)\;.
\end{aligned}
\end{equation}
From these two expressions, we can see that the weighted volume and surface do not satisfy the condition \eqref{eq:VrS}, so we cannot use the estimate \eqref{eq:sphsymest}. Instead, let use \eqref{eq:estcomp} and denote ${Q:=A_{\pm}/A_{0}}$ and ${\zeta:=\rho/A_{0}}$. The constants $A_{\pm}$ are given by the properties of the function ${\sinh ^2(\zeta ) \cosh (\zeta )-Q^2{\sinh ^2(\zeta  /Q)}}$. Since it is purely negative if ${Q<\sqrt{2/5}}$ and purely positive if ${Q>2/3}$, we may identify
\begin{equation}\label{eq:Apm0}
   A_{+}=\frac{2}{3}A_0\;, \quad A_{-}=\sqrt{\frac{2}{5}}A_0\;.
\end{equation}

\begin{figure}
    \centering
    \includegraphics[width=\columnwidth]{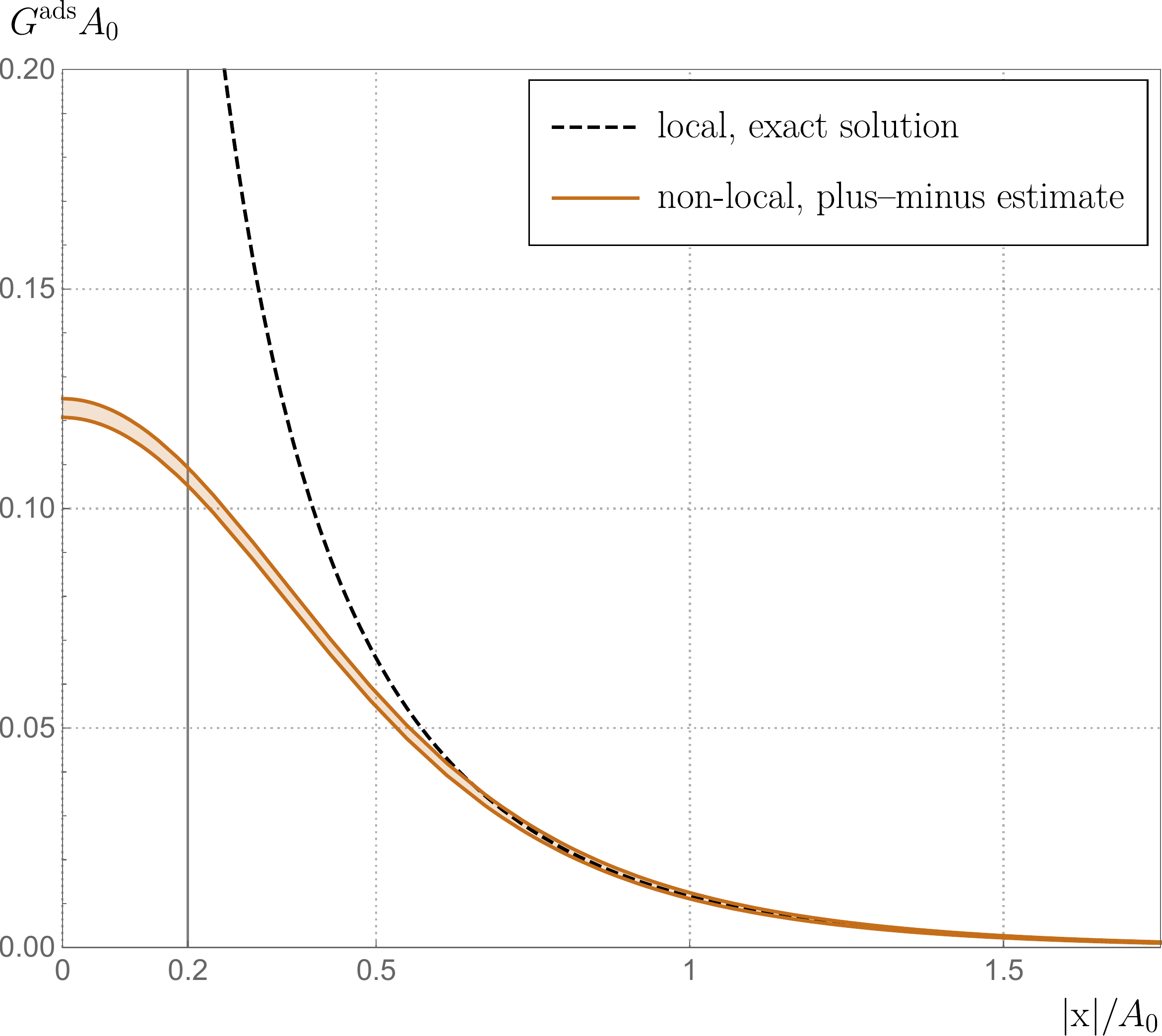}
    \caption{Static Green's functions centered at the origin calculated in anti-de Sitter spacetime. Dashed line is the exact local static Green's function given by \eqref{eq:locgfads}. Brown region describes the plus--minus estimate of the non-local static Green's function \eqref{eq:estcomp} with \eqref{eq:Apm0}. We set ${\ell/A_0=1/5}$.}
    \label{fig:ads}
\end{figure}

As before, for the comparison, we can find an exact expression for the local Green's function centered at the origin using \eqref{eq:Gsymloc}. Performing the integral, we arrive at the formula
\begin{equation}\label{eq:locgfads}
    G^{\textrm{ads}}_{\textrm{loc}}(\mathrm{x},\mathrm{o})=\frac{\csch\Big(\frac{|\mathrm{x}|}{A_0}\Big)+\gd\Big(\frac{|\mathrm{x}| }{A_0}\Big)-\frac{\pi}{2} }{16 \pi  A_0}\;,
\end{equation}
which reduces to the flat case ${1/4\pi|\mathrm{x}|}$ for ${A_0\to\infty}$ corresponding to the Minkowski limit of the anti-de Sitter spacetime. (Here $\gd$ denotes the Gudermannian function.) The above results are depicted in Fig.~\ref{fig:ads}.

%%%%%%%%%%%%%%%%%%%%%%%%%%%%%%%%%%%%%%%%%%%%%%%%%%%%%%%%%%%%%%%%%%%%%%%%%%%%%%%%
%%%%%%%%%%%%%%%%%%%%%%%%%%%%%%%%%%%%%%%%%%%%%%%%%%%%%%%%%%%%%%%%%%%%%%%%%%%%%%%%
%% Frequency-domain Green's functions

\section{Frequency-domain Green's functions} \label{sec:frequencydomaingreensfunctions}

Let us proceed to time-dependent scalar fields but in ultrastatic spacetimes. Solutions of non-local equations such as \eqref{eq:nlequation} with time-dependent sources are significantly more challenging to find and rather rare in the literature (see, e.g., \cite{Frolov:2015bia,Kolar:2021oba}). Moreover, it is well known that the spacetime Green's function of the non-local wave operator $-e^{-\ell^2\bar\square}\bar\square$ does not exists even for the Minkowski spacetime \cite{Buoninfante:2018mre,Heredia:2021pxp}. To overcome this issue we propose moving to the frequency domain and solving the non-local inhomomogeneous Helmholtz equation \eqref{eq:redeq2},
\begin{equation*}
    \boxed{-\big(\triangle+\omega^2\big)\hat{\phi}^{\omega}=e^{\ell^2\omega^2}e^{\ell^2\triangle}\hat{J}^{\omega}\;.}
\end{equation*}
Again, we employ the Green's function method.

\subsection{Frequency-domain Green's functions from heat kernels}

First, we consider the non-compact manifold. We define the \textit{frequency-domain Green's function} $\hat{\mathsf{G}}^{\omega}(\mathrm{x},\mathrm{y})$ as a solution of the above non-local equation with the source ${\hat{J}^{\omega}=\delta_{\mathrm{y}}}$ and the boundary condition ${\hat{\mathsf{G}}^{\omega}\to0}$ for ${D_{\mathrm{xy}}\to\infty}$ and ${\forall\omega\in\mathbb{R}}$. With the help of \eqref{eq:Kjeenadelta}, we may write
\begin{equation}\label{eq:GFfreq}
    -\big(\triangle+\omega^2\big)\hat{\mathsf{G}}^{\omega}(\cdot,\mathrm{y})=e^{\ell^2\omega^2}e^{\ell^2\triangle}\delta_{\mathrm{y}}=e^{\ell^2\omega^2}K_{\ell^2}(\cdot,\mathrm{y})\;.
\end{equation}
Since we work in the frequency domain, we allow for the complex solutions in general. However, it is clear that only the real part $\Re \hat{\mathsf{G}}$ should depend on the scale of non-locality $\ell$. The imaginary part $\Im \hat{\mathsf{G}}$ should be independent of $\ell$ because it comprises of a purely homogeneous solution. This (local) imaginary part can be chosen so that the character of the solutions is either retarded or advanced. The solution with an arbitrary source ${\hat{J}^{\omega}\in F {\mathcal{M}}}$ for each ${\omega\in\mathbb{R}}$ is then given by
\begin{equation}
    \hat{\phi}^{\omega}(\mathrm{x})=\langle \hat{\mathsf{G}}^{\omega}(\mathrm{x},\cdot),\hat{J}^{\omega}\rangle\;.
\end{equation}
By analogy with \eqref{eq:GGloc}, we may also introduce the local frequency-domain Green's function $\hat{\mathsf{G}}_{\textrm{loc}}^{\omega}$,
\begin{equation}\label{eq:hatphiloc}
    \hat{\mathsf{G}}^{\omega}= e^{\ell^2\omega^2}e^{\ell^2\triangle}\hat{\mathsf{G}}^{\omega}_{\textrm{loc}}\;, \quad -\big(\triangle+\omega^2\big)\hat{\mathsf{G}}^{\omega}_{\textrm{loc}}(\cdot,\mathrm{y})=\delta_{\mathrm{y}}\;.
\end{equation}

At this point, we would like to write an integral representation analogous to \eqref{eq:GreenusingKT}. Unfortunately, the na\"ive integral of $e^{\tau\omega^2}K_{\tau}$ could have serious problems with the convergence even in the local case (whenever ${\omega^2>\lambda_{\textrm{min}}}$). However, the analyticity in $\omega$ provides an interesting method of generating possible candidates for the frequency-domain Green's functions $\hat{\mathsf{G}}$: First, we replace ${\omega^2\to-z^2}$ in our non-local inhomogeneous Helmholtz equation turning it into the non-local screened Poisson equation. Then, we perform the integral
\begin{equation}\label{eq:theintegral}
    \int\limits_{\ell^2}^{\infty}\! d\tau\,e^{-\tau z^2}K_{\tau}(\mathrm{x},\mathrm{y})\;, \quad z>0\;.
\end{equation}
Finally, to get the formula for $\hat{\mathsf{G}}^{\omega}(\mathrm{x},\mathrm{y})$, we analytically continue the expression to ${z\to i\omega+0^+}$. Of course, the result obtained by this formal procedure must be always checked against definitions of the frequency-domain Green's functions \eqref{eq:GFfreq} and \eqref{eq:hatphiloc}. Also, let us remark that we have chosen the sign so that it leads to retarded solutions. (The advanced solutions are obtained by replacement ${\omega\to -\omega}$.) Upon taking the limit ${\omega\to0}$, we regain the formula for the static Green's function \eqref{eq:GreenusingKT}.

Let us move on to the compact manifolds. Notice that the field equations impose constraints of vanishing source for frequencies ${\omega^2=\lambda_{l}}$, i.e., ${\hat{J}_{l}^{\pm\sqrt{\lambda_l}}=\langle \hat{J}^{\pm\sqrt{\lambda_l}},\psi_{l}\rangle=0}$, see \cite{garfken67:math}. If ${\omega^2\neq\lambda_{k}}$, ${\forall k\in\mathbb{N}_0}$, then we define the Green's function with the source ${\hat{J}^{\omega}=\delta_{\mathrm{y}}}$,
\begin{equation}
\begin{aligned}
    -(\triangle+\omega^2)\hat{\mathsf{G}}^{\omega}(\cdot,\mathrm{y}) &=e^{\ell^2\omega^2}e^{\ell^2\triangle}\delta_{\mathrm{y}}=e^{\ell^2\omega^2}K_{\ell^2}(\cdot,\mathrm{y})
    \\
    &=\sum_{k=0}^{\infty}e^{-\ell^2(\lambda_k-\omega^2)}\psi_{k}(\cdot)\psi_{k}(\mathrm{y})\;.
\end{aligned}
\end{equation}
However, if ${\exists l\in\mathbb{N}_0}$: ${\omega^2=\lambda_{l}}$, then we have to modify the source in the definition of the frequency-domain Green's function to respect the restrictions described above, ${\hat{J}^{\pm\sqrt{\lambda_l}}=\delta_{\mathrm{y}}-\psi_l(\cdot)\psi_l(\mathrm{y})}$, 
\begin{equation}
\begin{aligned}
    -(\triangle+\lambda_l)\hat{\mathsf{G}}^{\pm\sqrt{\lambda_l}}(\cdot,\mathrm{y}) &=e^{\ell^2\lambda_l}e^{\ell^2\triangle}\big(\delta_{\mathrm{y}}-\psi_l(\cdot)\psi_l(\mathrm{y})\big) 
    \\
    &=e^{\ell^2\lambda_l}K_{\ell^2}(\cdot,\mathrm{y})-\psi_l(\cdot)\psi_l(\mathrm{y})
    \\
    &=\sum_{\mathclap{\substack{k=0 \\ k\neq l}}}^{\infty}e^{-\ell^2(\lambda_k-\lambda_l)}\psi_{k}(\cdot)\psi_{k}(\mathrm{y})\;.
\end{aligned}
\end{equation}
Here, we used the equations \eqref{eq:Kjeenadelta}, \eqref{eq:infsumeig}, and \eqref{eq:heatkerneleigen}. Also, notice that the indexing set of the sum is a consequence of ${\hat{J}_{l}^{\pm\sqrt{\lambda_l}}=0}$. The solution for an arbitrary source ${\hat{J}^{\omega}\in F\mathcal{M}}$, ${\omega\in\mathbb{R}}$, satisfying ${\hat{J}_{l}^{\pm\sqrt{\lambda_l}}=0}$, then still reads
\begin{equation}\label{eq:solutionphi}
    \hat{\phi}^{\omega}(\mathrm{x})=\langle \hat{\mathsf{G}}^{\omega}(\mathrm{x},\cdot),\hat{J}^{\omega}\rangle\;.
\end{equation}
Also, the relation to the local frequency-domain Green's function remains unchanged,
\begin{equation}
\begin{aligned}
    &\hat{\mathsf{G}}^{\omega} = e^{\ell^2\omega^2}e^{\ell^2\triangle}\hat{\mathsf{G}}^{\omega}_{\textrm{loc}}\;, 
    \\ 
    -\big(\triangle &+\omega^2\big)\hat{\mathsf{G}}^{\omega}_{\textrm{loc}}(\cdot,\mathrm{y}) =\begin{cases}\delta_{\mathrm{y}}\;,\phantom{{}-\psi_l(\cdot)\psi_l(\mathrm{y})} \;\; \omega^2\neq\lambda_k\;,
    \\
    \delta_{\mathrm{y}}-\psi_l(\cdot)\psi_l(\mathrm{y})\;, \;\; \omega^2 =\lambda_l\;.
    \end{cases}
\end{aligned}
\end{equation}

Similar to \eqref{eq:gfcompact}, we can also derive
\begin{equation}\label{eq:freqGFcompact}
\begin{aligned}
    \hat{\mathsf{G}}^{\omega} 
    &=\sum_{k=0}^{\infty}\frac{e^{-\ell^2(\lambda_k-\omega^2)}}{\lambda_k-\omega^2}\psi_{k}(\mathrm{x})\psi_{k}(\mathrm{y})\;, \;\;\omega^2\neq\lambda_k\;, 
    \\
    \hat{\mathsf{G}}^{\pm\sqrt{\lambda_l}}
    &=\sum_{\mathclap{\substack{k=0 \\ k\neq l}}}^{\infty}\frac{e^{-\ell^2(\lambda_k-\lambda_l)}}{\lambda_k-\lambda_l}\psi_{k}(\mathrm{x})\psi_{k}(\mathrm{y})+C_l\psi_{l}(\mathrm{x})\psi_{l}(\mathrm{y})\;,
\end{aligned}
\end{equation}
either from the corresponding integrals of $K_{\tau}$ or by the spectral expansion of the field equation. Here, $C_l$ are arbitrary constants that characterize the freedom in the homogeneous part of the solution due to ${-\triangle\psi_{l}=\lambda_l\psi_{l}}$. They will enable us not only to prescribe a certain value of the field at a given point but also to achieve a desired character of the solutions (retarded/advanced). Notice that the constants $C_l$ are absent for ${\omega^2\neq\lambda_k}$ because the equation admits no homogeneous solutions beyond the eigenfunctions. This also means that such a Green's function cannot produce retarded or advanced solutions. Let us also point out that the case ${\omega^2=\lambda_l}$ nicely reproduces the static Green's functions \eqref{eq:gfcompact} as a special subcase ${l=0}$ in contrast to the frequency-domain Green's functions with ${\omega^2\neq\lambda_k}$, which tend to blow up for ${\omega^2\to\lambda_k}$. Finally, it is not difficult to realize that the two formulas in \eqref{eq:freqGFcompact} are actually related through the limit (see \cite{Szmytkowski2007}, for the local case),
\begin{equation}
    \hat{\mathsf{G}}^{\pm\sqrt{\lambda_l}}\big|_{C_l=0}=\lim_{\omega^2\to\lambda_l}\frac{\partial}{\partial\omega^2}\left[\frac{\omega^2-\lambda_l}{e^{-\ell^2(\lambda_l-\omega^2)}}\hat{\mathsf{G}}^{\omega}\right]\;.
\end{equation}

\subsection{Example: Minkowski spacetime}

\begin{figure}
    \centering
    \includegraphics[width=\columnwidth]{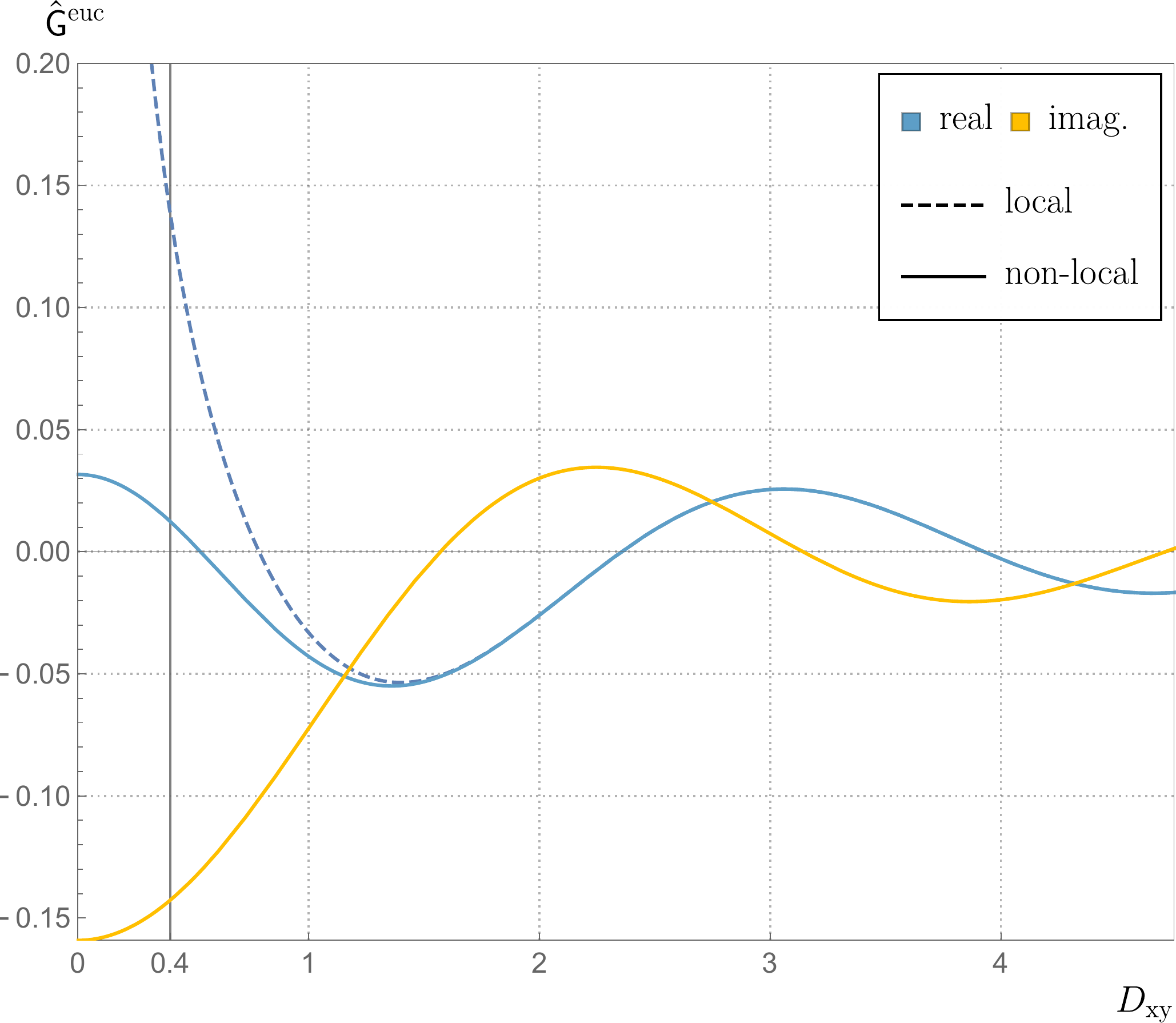}
    \caption{Frequency-domain Green's functions calculated in the Minkowski spacetime. Dashed and solid lines describe the local and non-local retarded frequency-domain Green's function, respectively, see \eqref{eq:FGFeuc} or \eqref{eq:reimGFeuc}. Blue color represents their real parts while the yellow color their imaginary parts. Dashed and solid yellow lines coincide because the local and non-local imaginary parts are identical. We set ${\ell=0.4\,\textrm{m}}$ and ${\omega=2\,\textrm{m}^{-1}}$.}
    \label{fig:euc}
\end{figure}

Let us start with the Minkowski spacetime. The frequency-domain Green's function can be obtained by performing the integral \eqref{eq:theintegral} with the heat kernel \eqref{eq:eucHK}. It leads to the integral that we previously denoted $H(z)$, ${z>0}$, in \eqref{eq:Hint}. This function can be analytically extended to all ${z\in\mathbb{C}}$. Evaluating $H(i\omega)$, we get the retarded frequency-domain Green's functions,
\begin{equation}\label{eq:FGFeuc}
    \hat{\mathsf{G}}^{\textrm{euc}} =H_{D_{\mathrm{xy}},\ell}(i\omega)\;,
    \quad
    \hat{\mathsf{G}}_{\textrm{loc}}^{\textrm{euc}} =H_{D_{\mathrm{xy}},0}(i\omega)\;.
\end{equation}
Of course, we should not just blindly rely on the analytic continuation of a diverging integral. However, the correctness of \eqref{eq:FGFeuc} can be easily verified by insertion to the left-hand side of \eqref{eq:GFfreq}. By realizing that the exponential and error functions commute with the complex conjugations, we can write real and imaginary parts of \eqref{eq:FGFeuc} explicitly as
\begin{equation}\label{eq:reimGFeuc}
\begin{aligned}
    \Re\hat{\mathsf{G}}^{\textrm{euc}} &= \frac{\Re\left[ e^{i\omega D_{\mathrm{xy}}}\erf\left(\tfrac{D_{\mathrm{xy}}}{2 \ell }+i\omega\ell \right)\right]}{4\pi D_{\mathrm{xy}}}\;, 
    \\ 
    \Re\hat{\mathsf{G}}^{\textrm{euc}}_{\textrm{loc}} &= \frac{\cos \left(\omega D_{\mathrm{xy}}\right)}{4\pi D_{\mathrm{xy}}}\;,
    \\
    \Im\hat{\mathsf{G}}^{\textrm{euc}} &=\Im\hat{\mathsf{G}}^{\textrm{euc}}_{\textrm{loc}}= -\frac{\sin \left(\omega D_{\mathrm{xy}}\right)}{4\pi D_{\mathrm{xy}}}\;.
\end{aligned}
\end{equation}
Here, we can clearly see that the imaginary part is independent of the scale of non-locality and bounded ${\forall\omega\in\mathbb{R}}$. By contrast, the real part blows up ${|\omega|\to\infty}$ in the non-local case. Because of this the spacetime Green's function of $-e^{-\ell^2\bar\square}\bar\square$, i.e., the inverse Fourier transform of $\hat{\mathsf{G}}^{\omega}/\sqrt{2\pi}$, does not exist in the non-local case. (In the local case, the spacetime Green's function of $-\bar\square$ is ${\delta(t-\tilde{t}-D_{\mathrm{xy}})/4\pi D_{\mathrm{xy}}}$.) The graphs of the obtained frequency-domain Green's functions are shown in Fig.~\ref{fig:euc}. One can check that the limit ${\omega\to0}$ reproduces the static Green's functions \eqref{eq:gfeuc}.

The frequency-domain Green's functions allow us to compute retarded solutions for various sources. The simplest time-dependent example, which was studied in \cite{frolovemitter2016} using a different method, is the point-like monochromatic emitter, ${J^t=J_0\cos(\Omega t)\delta_{\mathrm{y}}}$. Here, ${J_0}$ and $\Omega$ are two real positive constants. In the frequency domain, this source reads
\begin{equation}
    \hat{J}^{\omega}=\sqrt{\tfrac{\pi }{2}} J_0\left[ \delta (\omega -\Omega )+ \delta (\omega +\Omega )\right]\delta_{\mathrm{y}}\;.
\end{equation}
Employing \eqref{eq:solutionphi}, we can write the solution as
\begin{equation}
    \hat{\phi}^{\omega} =\sqrt{\tfrac{\pi }{2}} J_0 \hat{\mathsf{G}}^{\omega}\left[ \delta (\omega -\Omega )+ \delta (\omega +\Omega )\right]\;.
\end{equation}
In the time domain, thanks to the (anti-)symmetry in $\omega$, ${\Re\hat{\mathsf{G}}^{\omega}=\Re\hat{\mathsf{G}}^{-\omega}}$ and ${\Im\hat{\mathsf{G}}^{\omega}=-\Im\hat{\mathsf{G}}^{\omega}}$, it takes the form
\begin{equation}
   \phi^t=J_0\big[\Re\hat{\mathsf{G}}^{\Omega}\cos(\Omega t)-\Im\hat{\mathsf{G}}^{\Omega}\sin(\Omega t)\big]\;,
\end{equation}
which reduces to $J_0\cos[\Omega (D_{\mathrm{xy}}-t)]/4\pi D_{\mathrm{xy}}$ in the local case. In \cite{frolovemitter2016}, this result was derived through the 4-dimensional Fourier transform leading to the principal value integral in the form of the Hilbert transform. The heat kernel approach presented here is more a direct method. Unlike the 4-dimensional Fourier transform, it can be used also in the curved (ultrastatic) spacetimes.

\subsection{Example: Hyperbolic universe}

\begin{figure}
    \centering
    \includegraphics[width=\columnwidth]{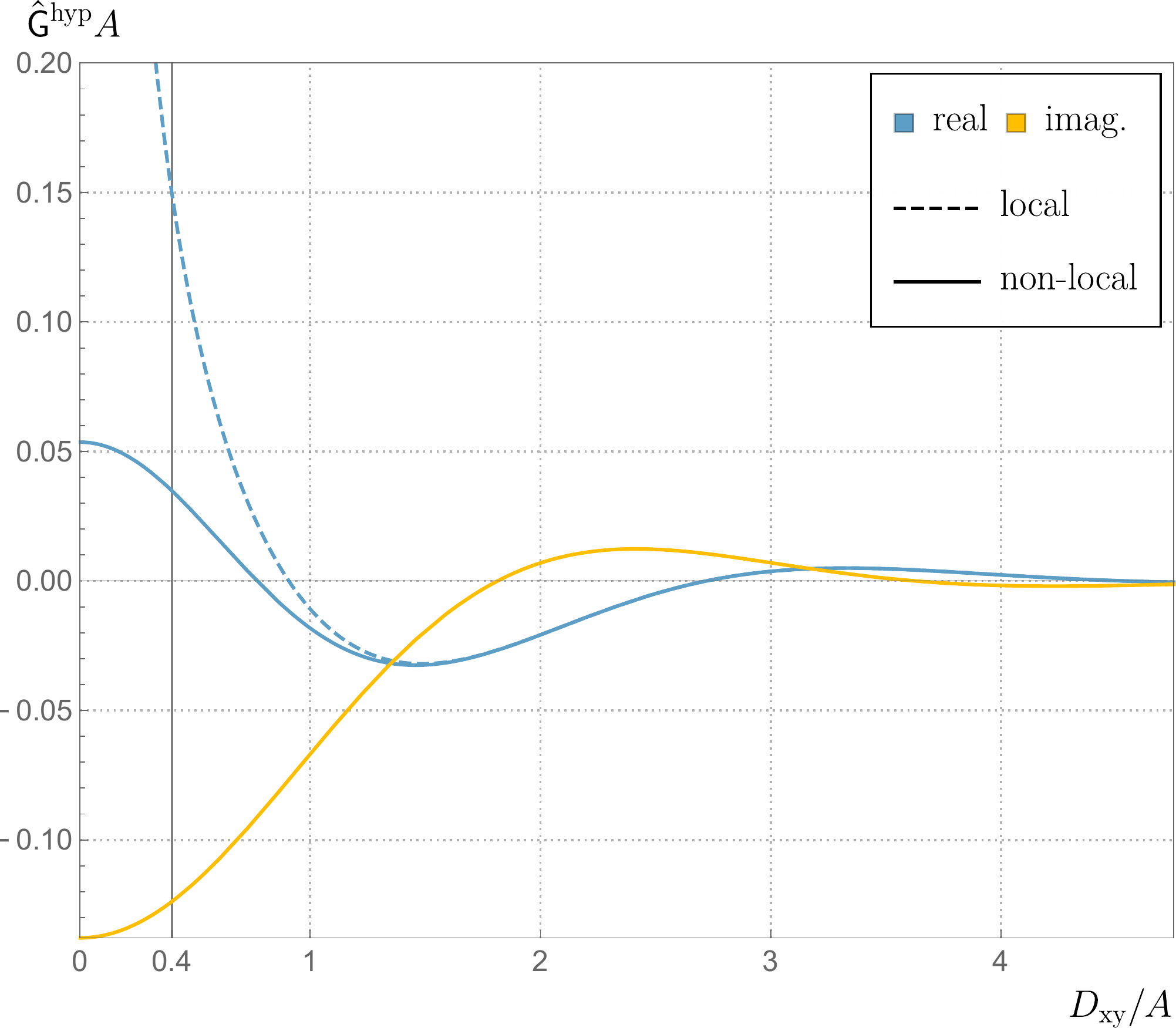}
    \caption{Frequency-domain Green's functions calculated in the hyperbolic universe for ${A|\omega|>1}$. Dashed and solid lines describe the local and non-local retarded frequency-domain Green's function, respectively, see \eqref{eq:FGFhyp} or \eqref{eq:reimGFhyp}. Blue color represents their real parts while the yellow color their imaginary parts. Dashed and solid yellow lines coincide because the local and non-local imaginary parts are identical. We set ${\ell/A=2/5}$ and ${A\omega=2}$.}
    \label{fig:hyp}
\end{figure}

Finding the frequency-domain Green's functions in the hyperbolic universe is very similar, since the integral \eqref{eq:theintegral} with the heat kernel \eqref{eq:Khyp} results in $H(\sqrt{z^2{+}1/A^2})$, ${z>0}$. Choosing the principal branch for the square root, we see that $\sqrt{z^2{+}1/A^2}$ has a branch cut at ${\Re{z}=0}$ and ${|\Im{z}|>1/A}$. Consequently, we can analytically continue the result to complex values with ${\Re z\geq0}$ and compute $H(\sqrt{(i\omega+0^{+})^2+1/A^2})$. This leads to the frequency-domain Green's functions,
\begin{equation}\label{eq:FGFhyp}
\begin{aligned}
    \hat{\mathsf{G}}^{\textrm{hyp}} &=\tfrac{{D_{\mathrm{xy}}}/{A}}{\sinh \left({D_{\mathrm{xy}}}/{A}\right)}H_{D_{\mathrm{xy}},\ell}\left(I/A\right)\;, 
    \\
    \hat{\mathsf{G}}^{\textrm{hyp}}_{\textrm{loc}} &=\tfrac{{D_{\mathrm{xy}}}/{A}}{\sinh \left({D_{\mathrm{xy}}}/{A}\right)}H_{D_{\mathrm{xy}},0}\left(I/A\right)\;,
    \\
    I &:=\sqrt{\big|A^2\omega^2-1\big|}\times\begin{cases}
    -i\;, & A\omega <-1\;, \\
    +1\;, & A|\omega| \leq 1\;, \\
    +i\;, & A\omega >1\;,
\end{cases}
\end{aligned}
\end{equation}
which can be verified again by direct insertion into \eqref{eq:GFfreq}. If ${A|\omega| \leq 1}$, then \eqref{eq:FGFhyp} are real and  qualitatively similar to the static Green's functions, i.e., the red lines in Fig.~\ref{fig:uni}. The absence of the imaginary part means that these functions describe standing waves. Further, if ${A|\omega| > 1}$, then the solutions have retarded character. The real and imaginary parts take the form,
\begin{equation}\label{eq:reimGFhyp}
\begin{aligned}
    \Re\hat{\mathsf{G}}^{\textrm{hyp}} &= \frac{\Re\left[ e^{\frac{I D_{\mathrm{xy}}}{A}}\erf\left(\tfrac{D_{\mathrm{xy}}}{2 \ell }+\frac{I \ell}{A}  \right)\right]}{4\pi A\sinh \left(\frac{D_{\mathrm{xy}}}{A}\right)}\;, 
    \\ 
    \Re\hat{\mathsf{G}}^{\textrm{hyp}}_{\textrm{loc}} &= \frac{\cos \left(\omega\sqrt{1-\frac{1}{A^2\omega^2}}D_{\mathrm{xy}}\right)}{4\pi A\sinh \left(\frac{D_{\mathrm{xy}}}{A}\right)}\;,
    \\
    \Im\hat{\mathsf{G}}^{\textrm{hyp}} &=\Im\hat{\mathsf{G}}^{\textrm{hyp}}_{\textrm{loc}}= -\frac{\sin \left(\omega\sqrt{1-\frac{1}{A^2\omega^2}} D_{\mathrm{xy}}\right)}{4\pi A\sinh \left(\frac{D_{\mathrm{xy}}}{A}\right)}\;.
\end{aligned}
\end{equation}
They have similar properties to the retarded frequency-domain Green's function in the Minkowski spacetime $\hat{\mathsf{G}}^{\textrm{euc}}$ (with an additional exponential damping), to which they reduce for ${A\to\infty}$. The real part blows up for ${|\omega|\to\infty}$ while the imaginary part remains bounded. These functions are depicted in Fig.~\ref{fig:hyp}.

\subsection{Example: Einstein universe}

We close this section with an example of the compact case. Considering the Einstein static spacetime, we can compute the frequency-domain Green's function by means of \eqref{eq:freqGFcompact}. Recall that $-\triangle$ on 3-sphere has a discrete spectrum with eigenvalues being the 3-dimensional spherical harmonics $\psi_{k,j}$ satisfying the addition theorem \eqref{eq:addtheo}. We arrive at the following results: If ${B^2\omega^2\neq k (k+2)}$, ${\forall k\in\mathbb{N}_0}$, then
\begin{equation}\label{eq:FGFsph1}
\begin{aligned}
    \hat{\mathsf{G}}^{\textrm{sph}} &=\sum_{k=0}^{\infty}\frac{e^{-\frac{\ell ^2}{B^2}\left({k (k+2) }-B^2\omega^2\right)}}{2 \pi ^2 B}\frac{(k+1)^2}{{k (k+2)}-B^2\omega ^2}
    \\
    &\feq\times\tfrac{\sin \left( (k+1)\frac{D_{\mathrm{xy}}}{B}\right)}{ (k+1) \sin \left(\frac{D_{\mathrm{xy}}}{B}\right)}\;,
    \\
    \hat{\mathsf{G}}^{\textrm{sph}}_{\textrm{loc}} &=\frac{ \U_{\sqrt{B^2\omega ^2+1}-1}\left(-\cos \left(\frac{D_{\mathrm{xy}}}{B}\right)\right)}{4 \pi B\sin \left(\pi  \sqrt{B^2\omega ^2+1}\right)}\;.
\end{aligned}
\end{equation}
Here, the letter $\U$ stands for the Chebyshev polynomial/function of the second kind. The solutions are real and describe standing waves. As we anticipated, they blow up for ${B^2\omega^2\to k (k+2)}$, ${k\in\mathbb{N}_0}$.

\begin{figure}
    \centering
    \includegraphics[width=\columnwidth]{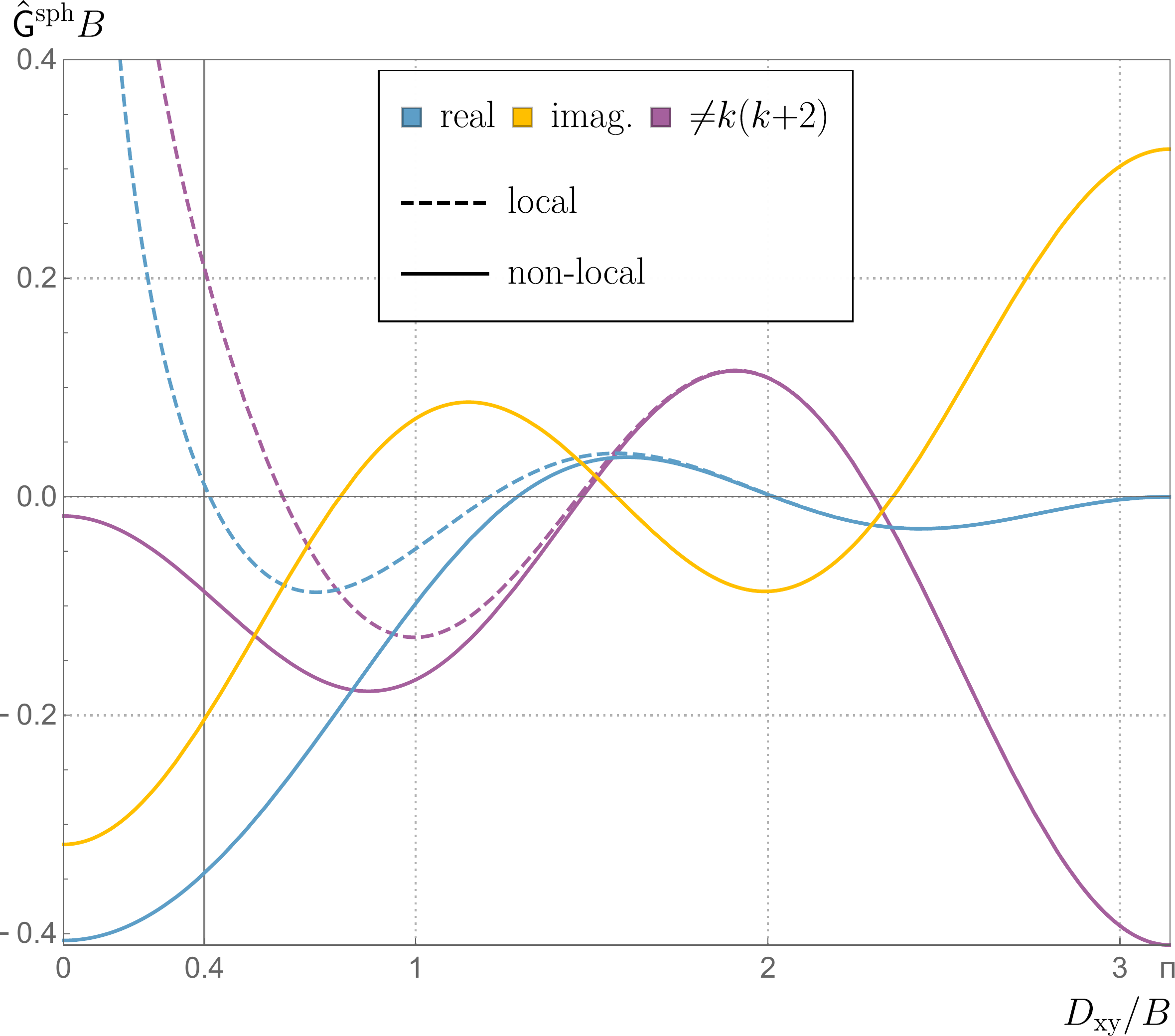}
    \caption{Frequency-domain Green's functions calculated in the Einstein universe. Dashed and solid lines describe the local and non-local case. Blue and yellow colors represent, respectively, the real and imaginary parts of the retarded frequency-domain Green's functions \eqref{eq:FGFsph2} with ${l=3}$. Dashed and solid yellow lines coincide because the local and non-local imaginary parts are identical. Purple color corresponds to the frequency-domain Green's functions \eqref{eq:FGFsph1} with ${B^2\omega^2=13\neq k (k+2)}$, ${\forall k\in\mathbb{N}_0}$, which are real (standing waves). We set ${\ell/B=2/5}$.}
    \label{fig:sph}
\end{figure}

On the other hand, if ${\exists l\in\mathbb{N}_0}$: ${B^2\omega^2=l (l+2)}$, then we find
\begin{equation}\label{eq:FGFsph2}
\begin{aligned}
    \Re\hat{\mathsf{G}}^{\textrm{sph}} &=\sum_{\mathclap{\substack{k=0 \\ k\neq l}}}^{\infty}\frac{e^{-\frac{\ell^2}{B^2}\left(k(k+2)-l(l+2)\right) }}{2 \pi^{2}B}\frac{(k+1)^2}{k(k+2)-l(l+2)}
    \\
    &\feq\times\left[\tfrac{\sin \left((k+1) \frac{D_{\mathrm{xy}}}{B}\right)}{(k+1)\sin\left( \frac{D_{\mathrm{xy}}}{B}\right)}-\tfrac{(-1)^{k-l}\sin \left((l+1) \frac{D_{\mathrm{xy}}}{B}\right)}{(l+1)\sin\left( \frac{D_{\mathrm{xy}}}{B}\right)}\right] \;, 
    \\
    \Re \hat{\mathsf{G}}^{\textrm{sph}}_{\textrm{loc}} &=\tfrac{\left(\pi -\frac{D_{\mathrm{xy}}}{B}\right) (l+1)\cot \left((l+1)\frac{D_{\mathrm{xy}}}{B}\right)+1}{4 \pi ^2 B }\tfrac{\sin \left((l+1) \frac{D_{\mathrm{xy}}}{B}\right)}{(l+1)\sin \left(\frac{D_{\mathrm{xy}}}{B}\right)}\;,
    \\
    \Im\hat{\mathsf{G}}^{\textrm{sph}} &=\Im \hat{\mathsf{G}}^{\textrm{sph}}_{\textrm{loc}}=-\sgn(B\omega)\tfrac{\sin \left((l+1)\frac{D_{\mathrm{xy}} }{B}\right)}{4 \pi  B \sin \left(\frac{D_{\mathrm{xy}}}{B}\right)}\;,
\end{aligned}
\end{equation}
where we have chosen the real parts of constants $C_l$ so that the real parts of the frequency-domain Green's functions vanish for the longest geodesics, ${D_{\mathrm{xy}}=\pi B}$. Furthermore, the imaginary part of $C_l$ have been set to achieve the retarded character (inspired by a formal replacement ${A\to i B}$ in its hyperbolic counterpart \eqref{eq:reimGFhyp}). Notice that the case ${l=0}$ corresponds to the static Green's function. Also, we recover $\hat{\mathsf{G}}^{\textrm{euc}}_{\textrm{loc}}$ when we take the flat limit ${B\to\infty}$ upon setting ${l=\sqrt{B^2\omega ^2+1}-1}$. Finally, let us mention that the local expressions in \eqref{eq:FGFsph1} and \eqref{eq:FGFsph2} are in agreement with the results of \cite{Szmytkowski2007} up to the homogeneous solutions. (Remark that $\Re \hat{\mathsf{G}}^{\textrm{sph}}_{\textrm{loc}}$ in \eqref{eq:FGFsph2} can be rewritten using the Chebyshev polynomials ${\T_l\left(\cos \left(x\right)\right)=\cos \left(lx\right)}$ and ${\U_l\left(\cos \left(x\right)\right)=\frac{\sin ((l+1) x)}{\sin x}}$.) The graphs of these Green's functions are plotted in Fig.~\ref{fig:sph}.

%%%%%%%%%%%%%%%%%%%%%%%%%%%%%%%%%%%%%%%%%%%%%%%%%%%%%%%%%%%%%%%%%%%%%%%%%%%%%%%%%%%%%%
%%%%%%%%%%%%%%%%%%%%%%%%%%%%%%%%%%%%%%%%%%%%%%%%%%%%%%%%%%%%%%%%%%%%%%%%%%%%%%%%%%%%%%
%% Conclusions

\section{Conclusions} \label{sec:conclusions}

In this paper we discussed solutions of the linear scalar field equation modified by the non-local exponential operator $e^{-\ell^2\square}$. We discussed two separate cases: i) static scalar field in static spacetimes and ii) time-dependent scalar field in ultrastatic spacetimes. Rewriting the problem from the 3-dimensional viewpoint, we showed that the equation takes the form of non-local Poisson/inhomogeneous Helmholtz equations in compact and non-compact weighted/Riemannian manifolds. In the first case, we found solutions by means of the exact and estimated static Green's functions, which can be derived from the heat kernels and their estimates. We also studied their regularity. In the second case, we introduced the frequency-domain Green's functions, can be related to the heat kernels through the analytic continuation. Finally, we demonstrated the general techniques discussed in this paper on several examples (ext. \& int. Schwarzschild, ultrastatic universes, anti-de Sitter).

Let us now go through possible extensions of our work and follow-up projects. The static and frequency-domain Green's functions we obtained here can be directly applied in the study of exact and estimated solutions generated by other (physically motivated) sources and also extended to more interesting static curved spacetimes. In the future works, we would like to elaborate more on the relation between the frequency-domain Green's function and the heat kernels to understand the analytic continuation better. In its current form, it only serves as a tool for generating possible candidates for the frequency-domain Green's functions. Another interesting direction of research would be the application of the presented methods to non-linear problems, such as the perturbative treatment of IDG or SFT/PST with non-linear potentials. Finally, we would like to extend our results also to higher-order exponential operators such as $e^{(-\ell^2\square)^N}$, ${N\in\mathbb{N}}$, perhaps with the help of the recent work \cite{Barvinsky:2019spa}. As discussed in \cite{frolovemitter2016,Kolar:2021oba}, the operators with even powers of $\Box$ are expected to remove issues with divergences for high frequencies that we also observed in the frequency-domain Green's functions.

%%%%%%%%%%%%%%%%%%%%%%%%%%%%%%%%%%%%%%%%%%%%%%%%%%%%%%%%%%%%%%%%%%%%%%%%%%%%%%%%%%%%%%
%%%%%%%%%%%%%%%%%%%%%%%%%%%%%%%%%%%%%%%%%%%%%%%%%%%%%%%%%%%%%%%%%%%%%%%%%%%%%%%%%%%%%%
%% Acknowledgements

\section*{Acknowledgements}

The author would like to thank to Carlos Heredia Pimienta (Barcelona, Spain) and Anupam Mazumdar (Groningen, Netherlands) for stimulating discussions. This work was supported by Netherlands Organization for Scientific Research (NWO) grant no. 680-91-119.

%%%%%%%%%%%%%%%%%%%%%%%%%%%%%%%%%%%%%%%%%%%%%%%%%%%%%%%%%%%%%%%%%%%%%%%%%%
%%%%%%%%%%%%%%%%%%%%%%%%%%%%%%%%%%%%%%%%%%%%%%%%%%%%%%%%%%%%%%%%%%%%%%%%%%%%%%%%%%%%%%
%% REFERENCES

%\bibliographystyle{apsrev4-2}
%\apptocmd{\thebibliography}{\raggedright}{}{}

%\clearpage

%\vfill

\bibliography{references.bib}

\end{document}